\documentclass[
aps,
prx,
amsmath,
amssymb,
reprint,
longbibliography,
]{revtex4-2}

\usepackage[utf8]{inputenc}
\usepackage{graphicx}
\usepackage[dvipsnames,table]{xcolor}
\usepackage{braket}
\usepackage[percent]{overpic}
\usepackage{amssymb}
\usepackage{pict2e}
\usepackage{hyperref}
\usepackage{braket}
\usepackage{subcaption}
\usepackage{float}
\usepackage{placeins}
\newcommand{\toadd}[1]{{#1}}

\begin{document}
\title{Excitation Amplitude Sampling for Low Variance Electronic Structure on Quantum Computers}
\author{Connor Lenihan$^1$}
\author{Oliver J. Backhouse$^1$}%
\author{Basil Ibrahim$^1$}
\author{Tom W. A. Montgomery$^2$}
\author{Phalgun Lolur$^2$}
\author{M. J. Bhaseen$^1$}
\author{George~H.~Booth$^1$}%
\email{george.booth@kcl.ac.uk}
\affiliation{$^1$Department of Physics, King's College London, Strand, London WC2R 2LS, U.K.}
\affiliation{$^2$Capgemini Quantum Lab, Capgemini, 147-151 Quai du Pr\'esident Roosevelt, 92130 - Ile-de-France, Issy-les-Moulineaux, France}


\begin{abstract}
    We combine classical heuristics with partial shadow tomography to enable efficient protocols for extracting information from correlated {\em ab initio} electronic systems encoded on quantum devices. By proposing the use of a correlation energy functional and sampling of a polynomial set of excitation amplitudes of the quantum state, we can demonstrate an almost two order of magnitude reduction in required number of shots for a given statistical error in the energy estimate, as well as observing a linear scaling to accessible system sizes. Furthermore, we find a high-degree of noise resilience of these estimators on real quantum devices, with up to an order of magnitude increase in the tolerated noise compared to traditional techniques. While these approaches are expected to break down asymptotically, we find strong evidence that these large system arguments do not prevent algorithmic advantage from these simple protocols in many systems of interest. We further extend this to consider the extraction of beyond-energetic properties by mapping to a coupled cluster surrogate model, as well as a natural combination within a quantum embedding framework. This embedding framework avoids the unstable self-consistent requirements of previous approaches, enabling application of quantum solvers to realistic correlated materials science, where we demonstrate the volume-dependence of the spin gap of Nickel Oxide.
\end{abstract}
\maketitle

\color{black}

\section{Introduction} \label{sec:intro}
The simulation of physical quantum systems is one of the key applications driving the development of digital quantum devices, and an original motivation for their inception~\cite{Feynman1982,doi:10.1021/acs.chemrev.9b00829,doi:10.1021/acs.chemrev.8b00803,RevModPhys.92.015003}. While there has been a huge growth of algorithms to leverage the potential for quantum information processing, a significant challenge limiting their application to realistic systems is the extraction of observables. These observables are cast as the expectation values of samples from a probability distribution via measurements of the quantum device, and the overhead and associated variance with these repeated measurements significantly limits the scale and scope of the information which can be extracted from the simulated system. 

The most important observable is the energy of a quantum state. This can be estimated via a digital readout when quantum phase estimation (QPE) and related algorithms act on a (near) eigenstate of the system Hamiltonian~\cite{kitaev1995quantummeasurementsabelianstabilizer,Kivlichan2020improvedfault,PRXQuantum.3.010318,PRXQuantum.4.040341}, or via a direct sampling of the expectation value. This direct sampling approach relies on the low-body nature of {\em ab initio} interactions in a quantum system, to decompose the Hamiltonian into a weighted average of a polynomial (yet large) number of Pauli strings. These can be estimated and averaged over samples before combination to a statistical estimate of the energy. This is a core primitive routine at the heart of current and near-term `NISQ' algorithms, which trade circuit depth and qubits for increased circuit repetitions and measurements (shots)~\cite{Preskill2018quantumcomputingin,McClean_2016,TILLY20221,BLEKOS20241}. However, it was realized early on that for realistic systems, the $\mathcal{O}[N^4]$ native Pauli strings of the {\em ab initio} Hamiltonian would result in a `variance problem'~\cite{PhysRevA.92.042303}. This is due to the large number of shots required, $M$, to reach the low statistical uncertainty in the energy estimate required for chemical and materials science applications. Intense research has therefore taken place in recent years to reduce the number of shots for an associated variance of the estimate, based on grouping the Pauli strings into commuting cliques~\cite{C8SC05592K,jena2019paulipartitioningrespectgate,Crawford2021efficientquantum,PhysRevA.101.062322,9248636,doi:10.1021/acs.jctc.0c00008,10.1063/1.5141458,Wu2023overlappedgrouping,Yen2023,PRXQuantum.2.040320,PhysRevX.10.031064,Burns_2025}, using randomized approaches~\cite{Huang_classicalShadows,Elben2023,Hadfield2022,dutt2023practicalbenchmarkingrandomizedmeasurement,PhysRevLett.127.110504,bian2025adaptivedepthrandomizedmeasurementfermionic}, adaptive strategies~\cite{Shlosberg2023adaptiveestimation,PhysRevA.109.062404,doi:10.1021/acs.jctc.3c01113,gresch2025reducingsamplingcomplexityenergy,Gresch2025}, factorized operators~\cite{Huggins2021}, or leveraging machine learned or signal processing models~\cite{10.1063/5.0219663,Iouchtchenko_2023,bassi2025noisyobservablesaccurateground}.

For an additive error in the energy estimate, $\epsilon$, early forecasts for a minimal-basis 14-qubit water molecule, resulted in an estimate of $M\sim 10^9$ for a modest $\epsilon=1$mE$_\mathrm{h}$ statistical error, despite this system being simple to solve exactly on classical resources~\cite{PhysRevA.92.042303}. More recent state of the art variance reduction methods have reduced this estimate to $M\sim 10^8$ shots~\cite{Huggins2021,10.1063/5.0173591}, or $M\sim 10^7$ for the slightly higher target error of 1.6mE$_\mathrm{h}$ (taken as chemical accuracy)~\cite{Gresch2025}. For a more challenging (albeit still relatively small) Ferredoxin target system (Fe$_2$O$_2$) in 56 orbitals, an estimate of $10^{13}$ shots was given per single energy estimate~\cite{PhysRevA.92.042303}. While estimates of long-term circuit execution times vary with both hardware and details of the circuit and error mitigation or correction overheads~\cite{PhysRevApplied.17.044016,PRXQuantum.2.030305}, these 
required shot numbers and their scaling with system size constitute a severe limitation in the scope of systems that can be considered~\cite{doi:10.1021/acs.jpca.2c04726,doi:10.1021/acs.jctc.3c00851,PhysRevLett.130.100801}. 

This is often considered as motivation for QPE-like algorithms with a digital readout of the energy. These are expected to come to the fore once fault tolerance of the devices permits the depth of circuits required, which will scale linearly with the inverse precision, in a Heisenberg-limited fashion as $\sim1/\epsilon$~\cite{baysmidt2025faulttolerantquantumsimulationgeneralized}. This allows for an energy estimate which avoids an explicit dependence on a number of shots (although there will likely still be a dependence from the initial state preparation problem~\cite{Lee2023}). However, this does not entirely solve the problem, since the extraction of other critically important expectation values for characterizing electronic structure, such as electrical moments, reduced density matrices, nuclear forces and other response properties are not able to be extracted in this fashion. Returning to a direct sampling of these associated operators could still be required~\cite{PhysRevResearch.4.043210,PhysRevA.104.032405}.

As a final motivating challenge for the developments of this work, we consider the importance of coupling quantum resources to the description of a wider system of interest on classical resources, enabling multi-scale and multi-resolution hybrid quantum-classical workflows. For application of even long-term quantum algorithms to realistic chemical and materials science applications, these hybrid approaches are likely to be essential, allowing for the targeted application of precious quantum resources to subspaces of systems featuring particularly strong correlations~\cite{ALEXEEV2024666,Ma2020,Vorwerk2022}. This is achieved with `quantum embedding' methods, which require the extraction of a compact electronic variable from the quantum solver of the subspace, in order to couple it to the wider classical description of the system~\cite{doi:10.1021/acs.accounts.6b00356}. These variables could be the reduced density matrix of the subspace in the case of density matrix embedding theory or active space quantum chemical methods~\cite{rubin2016hybridclassicalquantumapproachlargescale, Vorwerk2022,DMET_montanaro,D2SC01492K,Cao2023,PhysRevResearch.3.013184,Gujarati2023,PhysRevResearch.3.033230}, the spectral moments~\cite{PhysRevResearch.3.033230,PhysRevResearch.6.023110}, or the subspace Green's function in dynamical mean-field theory~\cite{PhysRevX.6.031045,rungger2020dynamicalmeanfieldtheory,selisko2024dynamicalmeanfieldtheory,Keen_2020,PhysRevB.107.165155,Jaderberg_2020,jamet2024andersonimpuritysolverintegrating}, all of which have been investigated as multi-scale quantum-classical workflows for electronic structure. However, the task of efficiently sampling and extracting faithful representations of these quantum variables remains a key challenge, with noisy self-consistent loops between the quantum and classical descriptions compounding the difficulty in application of these approaches~\cite{Keen_2020}.

In this work we approach these fundamental challenges via a unified framework of sampling polynomially scaling sets of configurational amplitudes. Configurational sampling on quantum devices has been enabled in recent years by the rapid growth of shadow tomography for efficient estimation of probability amplitudes~\cite{aaronson2018shadow,Huang_classicalShadows,Elben2023,PhysRevLett.127.110504}, finding application in electronic structure for the sampling of reduced density matrices~\cite{PhysRevLett.127.110504,bian2025adaptivedepthrandomizedmeasurementfermionic}, quantum-enhanced Monte Carlo techniques~\cite{Huggins_unbiasingAFQMC,Wan2023_matchgate_shadows,PhysRevResearch.6.043063,Blunt_fciqmc_w_shadows,Kiser_2024}, quantum-tailored coupled cluster~\cite{doi:10.1021/acs.jctc.4c00037,PhysRevResearch.6.023230}, and estimation of spectral gaps~\cite{PRXQuantum.6.010352} among others~\cite{PhysRevResearch.4.033173,Hwang_2025}. In these approaches, randomized samples of a quantum state are drawn to provide a compact joint estimate of a set of operators which can subsequently be used to enhance a classical algorithm or enable extraction of expectation values of interest. 
Here, we broaden the scope by leveraging the combination of shadow tomographic techniques with classical heuristic states in electronic structure, to develop a simple framework to enhance the extraction of total energies of a quantum state, beyond energetic properties, and describe a more robust framework for multi-scale quantum embedding protocols for large systems.
The combination of quantum algorithms with classically derived information and state constraints, as we consider here, can often formally introduce an implicit exponential scaling with system size. This scaling appears frequently in quantum algorithms, including in state preparation protocols or error mitigation techniques~\cite{Lee2023,Takagi2022,Quek2024,zou2025multireferenceerrormitigationquantum}. 
However, there is increasing evidence that often these asymptotic exponential scaling arguments do not preclude practical benefits in highly relevant application areas and can still be used for algorithmic quantum advantage for systems out of reach of classical approaches~\cite{berry2024rapidinitialstatepreparation,D4FD00141A,tubman2018postponingorthogonalitycatastropheefficient,doi:10.1021/acscentsci.8b00788,zimboras2025mythsquantumcomputationfault,aharonov2025importanceerrormitigationquantum}.

In Sec.~\ref{sec:energy} we describe the extraction of the total energy for a given quantum state of an {\em ab initio} Hamiltonian, utilizing a simple mean-field classical surrogate and polynomial configurational shadow sampling. By switching to a non-variational {\em correlation} energy functional, the quality of the results now depend on the accuracy of the classical heuristic used, but nevertheless can often result in a significantly reduced variance for a given shot budget, and up to two orders of magnitude reduction in circuit repetitions even for simple mean-field heuristics.
We also show a scaling with system size which is favorable well into a practically relevant regime compared to direct sampling of the energy traditionally employed on current generation quantum devices. Furthermore, we demonstrate the resilience of the estimate with respect to noise in the circuit, with real superconducting quantum devices used to validate numerical simulated experiments. 

In Sec.~\ref{sec:properties} we extend this to the extraction of properties beyond the total energy of the quantum state by relying on a coupled cluster surrogate model, showing a quantum enhancement of long-range anti-ferromagnetic spin correlation functions in hydrogen chains. Finally, in Sec.~\ref{sec:embedding}, we show how these approaches can be naturally combined with a quantum embedding to offload small fragments of larger systems to a quantum device for solution, and enable efficient reconstruction of these solutions for the total energy of extended systems. The variable which couples the fragment description on the quantum computer to the wider classical description of its environment in this case are the wave function amplitudes themselves, leading to a simplified approach and automatic enforcement of many physical constraints on the full system state. These include global particle number and thus avoid the requirement of self-consistent optimization of e.g. chemical potentials of the fragments which is a severe practical limitation of many other quantum embedding approaches. This approach is demonstrated for a fully {\em ab initio} calculation of Nickel Oxide, a paradigmatic correlated material, computing the spin gap and equation of state for this system in a fashion which overcomes many of the drawbacks of existing embedding techniques when combined with quantum solvers. 

\section{Energy estimates} \label{sec:energy}
\subsection{Mixed energy functional}

We start by assuming that we have the exact ground eigenstate, $|\Psi\rangle$, of an {\em ab initio} Hamiltonian, $\mathcal{H}$, encoded on a qubit register. We can represent this Hamiltonian in an arbitrary basis of single-particle states as
\begin{equation}
    \mathcal{H}=\sum_{pq} h_{pq} {\hat c}_p^\dagger {\hat c}_q+ \frac{1}{2}\sum_{pqrs} v_{pqrs} {\hat c}_p^\dagger {\hat c}_q^\dagger {\hat c}_s {\hat c}_r + E_\text{nuc} ,  \label{eq:Ham} 
\end{equation}
where $h_{pq}$ and $v_{pqrs}=\langle pq|rs \rangle$ are the one- and two-body terms in the Hamiltonian, and $E_\text{nuc}$ is the scalar nuclear repulsion energy. This Hamiltonian contains a number of individual second quantized operators which scales as $\mathcal{O}[N^4]$, where $N$ is the number of degrees of freedom (orbitals). A direct sampling estimate of the energy first decomposes these operators into a weighted sum of $\mathcal{P}$ individual Pauli strings, ${\hat P}_a \in \{\mathbb{I}, X, Y, Z\}^{\otimes N}$. This can be achieved in a number of different encoding strategies~\cite{TILLY20221}, to give
\begin{equation}
    \mathcal{H}=\sum_a^{\mathcal{P}} w_a {\hat P}_a ,
\end{equation}
where $w_a$ are the weights. The energy can then be computed as
\begin{equation}
E = \sum_a^{\mathcal{P}} w_a \langle \Psi | {\hat P}_a | \Psi \rangle . \label{eq:pureestimate}
\end{equation}
Each ${\hat P}_a$ can be diagonalized with a Clifford circuit and averaged over computational basis measurements to accumulate the Pauli expectation values independently.
Since $|\Psi\rangle$ is not an eigenstate in general of any ${\hat P}_a$, there is an associated (unknown) variance in sampling each Pauli string expectation value, denoted $\mathrm{Var}[P_a]=1-\langle \Psi | {\hat P}_a | \Psi \rangle^2$. Since $\mathcal{P} \propto N^4$ in {\em ab initio} electronic structure, an analysis of the number of shots required, $M$, for a given random error, $\epsilon$, in the overall energy estimate yields~\cite{TILLY20221}
\begin{equation}
    M \propto \frac{N^4}{\epsilon^2} \sum_a^\mathcal{P} w_a^2 \text{Var}[{\hat P}_a] . \label{eq:shotspure}
\end{equation}
This simplistic analysis assumes a uniform shot allocation across all Pauli terms (which would be more ideally distributed according to the (unknown) distribution $\sim|w_a|\sqrt{\text{Var}[{\hat P}_a]}$), and does not account for the fact that the operators can be grouped into mutually commuting sets and jointly measured, both of which are common starting points for variance reduction techniques~\cite{C8SC05592K,jena2019paulipartitioningrespectgate,Crawford2021efficientquantum,PhysRevA.101.062322,9248636,doi:10.1021/acs.jctc.0c00008,10.1063/1.5141458,Wu2023overlappedgrouping,Yen2023,PRXQuantum.2.040320,PhysRevX.10.031064,Burns_2025,Shlosberg2023adaptiveestimation,PhysRevA.109.062404,doi:10.1021/acs.jctc.3c01113,gresch2025reducingsamplingcomplexityenergy,Gresch2025,10.1063/5.0219663,Iouchtchenko_2023}. However, this simple analysis is sufficient for our purposes, and demonstrates the `variance problem' in this estimator.

We now consider an alternative linear energy functional of the statevector $|\Psi \rangle$, as
\begin{equation}
    E = \frac{\langle \Phi | \mathcal{H} | \Psi \rangle}{\langle \Phi | \Psi \rangle} , \label{eq:mixedestimator}
\end{equation}
where $|\Phi\rangle$ is denoted a `trial' state. This energy estimator is exact and equal to that of Eq.~\ref{eq:pureestimate} in the case that {\em either} $|\Psi \rangle$ or $| \Phi \rangle$ is the eigenstate of interest, and that $\langle \Phi | \Psi \rangle \neq 0$. We term the estimator in Eq.~\ref{eq:mixedestimator} the `mixed' estimator, in contrast to the `Pauli' estimator of Eq.~\ref{eq:pureestimate}, noting that it is not a variational functional for approximate states. This energy estimator is widely used across electronic structure theory, from perturbative expansions and coupled-cluster theory~\cite{RevModPhys.79.291} to projector Monte Carlo methods~\cite{10.1063/5.0232424,10.1063/1.3193710}, where it can be efficiently evaluated.

We can separate the state of interest, $|\Psi\rangle$, into a part which is collinear with the trial state, and an orthogonal correction to this, as
\begin{equation}
    |\Psi \rangle = c_0 |\Phi \rangle + c_\perp | \Psi_\perp \rangle ,
\end{equation}
where $c_0=\langle \Phi | \Psi \rangle$, $|\Psi_\perp \rangle$ is the normalized component of $|\Psi\rangle$ which spans the orthogonal complement to $|\Phi\rangle$, and $c_\perp=\langle \Psi_\perp | \Psi\rangle = \sqrt{1-c_0^2}$. Inserting this decomposition into Eq.~\ref{eq:mixedestimator} gives
\begin{align}
    E &= \frac{\langle \Phi | \mathcal{H} (c_0 | \Phi \rangle + c_\perp | \Psi_\perp \rangle )}{\langle \Phi | \Psi \rangle} , \\
    &= \langle \Phi | \mathcal{H} | \Phi \rangle + \frac{c_\perp}{c_0} \langle \Phi | \mathcal{H} | \Psi_\perp \rangle . \label{eq:mixedexplicit}
\end{align}
This therefore decomposes the total energy of this mixed estimator into a total energy of the trial state and a correction which couples the state of interest to the trial. Importantly, the correction term to the trial energy doesn't require knowledge of the whole state $|\Psi\rangle$, since it is only coupled to the first-order interacting space of the trial state, $|\Phi\rangle$. Put another way, since $\mathcal{H}$ is a sum of (at most) two-body operators, it cannot couple $|\Phi \rangle$ to the entire Hilbert space in which $|\Psi\rangle$ is represented. Here, we exploit this such that the component of $|\Psi_\perp \rangle$ which couples to $|\Phi \rangle$ via the Hamiltonian can be obtained efficiently, to allow use of this alternate estimator.

A simple example of a trial state for which this can be performed is a single mean-field (e.g. Hartree--Fock) Slater determinant state, which we denote by $|\phi\rangle$ and consider as the trial state for the rest of this section. This state is represented by a single configurational basis state if the orbitals of Eq.~\ref{eq:Ham} are canonicalized (i.e. eigenfunctions of the corresponding mean-field Hamiltonian). Alternatively, this state can be generated simply by the action of a fermionic Gaussian unitary circuit on a configuration, introducing single-body orbital rotations via $N$ layers of Givens rotations and a single layer of phase gates~\cite{PhysRevApplied.9.044036}. Within a canonical representation, we label hole states as $i, j, \dots$ and particle states as $a, b, \dots$, allowing a rewriting of the full statevector in the particle number conserving sector of the Hilbert space as
\begin{equation}
    \ket{\Psi} = c_0\ket{\phi} + \sum_{ia} c_{i}^a \ket{\phi_i^a} + \sum_{ijab} c_{ij}^{ab} \ket{\phi_{ij}^{ab}} + ..., \label{eq:excitvec}
\end{equation}
where $\ket{\phi_i^a}$ represents the trial mean-field state with an electron removed from the orbital $i$ (occupied in the Hartree-Fock state) and excited to the unoccupied orbital $a$, with double and higher excitations represented similarly as $\ket{\phi_{ij...}^{ab...}}$, and continuing until all electrons are promoted to particle states. This conservation of particle number via consideration of the excitation amplitudes from the Fermi vacuum truncates the full set of bit strings in the statevector to those with a Hamming weight given by the number of electrons.

Crucially, the first-order interacting space of $|\phi \rangle$ that couples the trial to $|\Psi \rangle$ (as represented in Eq.~\ref{eq:excitvec}) truncates after the two-body excitations, due to the fact that the Hamiltonian of Eq.~\ref{eq:Ham} is only two-body. Therefore, the energy of Eq.~\ref{eq:mixedexplicit} can be found from the mean-field energy and a knowledge of just the (polynomial number of) $c_i^a$ and $c_{ij}^{ab}$ excitation amplitudes. A further small simplification comes if a Hartree--Fock state is used, in which case the single excitation amplitudes ($c_i^a$) are also zero due to Brillouin's theorem. This leads to a compact energy expression which can be computed efficiently on classical resources if the $c_{ij}^{ab}$ amplitudes can be found, given by
\begin{equation}
    E = E_{\text{HF}} + \frac{1}{c_0} \sum_{ij}\sum_{ab}c_{ij}^{ab} (2v_{ijab}-v_{ijba}), \label{eq:pert_energy}
\end{equation}
where $E_{\text{HF}}$ is the Hartree--Fock total energy. This form is equivalent to perturbative expressions for energy corrections of mean-field states~\cite{PhysRev.46.618}. The expression only relies on the classically computed mean-field energy, and the two-electron Coulomb repulsion integrals that couple the particle and hole channels in the Hamiltonian, not the one-body matrix elements, $h_{pq}$, of Eq.~\ref{eq:Ham} derived from the kinetic energy or nuclear repulsion contributions.

The mixed estimator of Eq.~\ref{eq:mixedestimator} can therefore be efficiently computed from Eq.~\ref{eq:pert_energy} in the case of a mean-field trial state if a limited polynomial-sized set of configurational amplitudes, $c_0$ and $c_{ij}^{ab}$, can be tomographically extracted in an efficient manner from the state of interest on the quantum computer. This could be achieved via individual Hadamard tests or related approaches for these overlaps~\cite{tazi2024shallowquantumscalarproducts}, as considered in e.g. Ref.~\cite{PhysRevResearch.4.033173}. However a more effective strategy is to use the classical shadows approach, as suggested by Huggins {\em et al.} in their quantum-enhanced QMC approach~\cite{Huggins_unbiasingAFQMC}. We will consider this use of classical shadows in this context, before returning to the question of the properties of this mixed estimator for the total energy, the expected variance with number of shots, and numerical experiments on simulated and real hardware.

\subsection{Configurational sampling via classical shadows} \label{sec:shadowsampling}

Classical shadow tomography has emerged over the last few years as a protocol for measuring expectation values over a set of multiple (or even unknown) operators of a quantum state~\cite{aaronson2018shadow, Huang_classicalShadows}. Depending on the operators in the set and the specifics of the protocol, this scheme can exhibit as little as logarithmic scaling in the required measurements with the number of jointly sampled expectation values. 
Specifically, with high probability we can bound the number of shots, $M$, required to reach an additive precision of $\epsilon$ in all operators that we wish to measure, $\{\hat{O}_j\}$, as
\begin{equation}
    M \geq \frac{\log(L)}{\epsilon^2} \max_j || \hat{O}_j ||^2_{\text{shadow}} , \label{eq:shadowshots}
\end{equation}
where the shadow norm $||.||_{\text{shadow}}$ is dependent on the type of measurements taken as part of the protocol \cite{Huang_classicalShadows} and $L$ is the number of operators in the set $\{\hat{O}_j\}$ we aim to estimate, all with a precision better than $\epsilon$.

The approach first selects a unitary operator, $U_k$, randomly and with uniform probability from an ensemble, $\mathcal{U}$, and performs a measurement in the computational basis to obtain a bitstring $|b_k\rangle$. This process is repeated $M$ times, with a compact representation of the selected unitary, $U_k$, and measured bitstring, $|b_k \rangle$, stored for each repetition, for classical postprocessing later. The expectation of this process over many samples is a linear map on the underlying density operator of the state, $\rho=| \Psi \rangle \langle \Psi |$, as
\begin{equation}
    \mathcal{M}(\rho) = \mathbb{E}_k[U_k^\dagger |b_k \rangle \langle b_k|U_k] . \label{eq:shadowdensityoperator}
\end{equation}
This pseudo-channel can be analytically inverted for certain ensembles of unitaries, $\mathcal{U}$. In particular, in this work we select $\mathcal{U}$ to be the ensemble of operators over the $N$-qubit Clifford group, $\mathcal{C}_N$, \toadd{whose unitary operators can be decomposed into primitive Clifford gates with a depth which grows only logarithmically with $N$, even without all-to-all connectivity~\cite{doi:10.1126/science.adv8590,PhysRevLett.133.020602}. This is likely far shallower than any electronic structure state preparation circuits, while further optimization and approximation of these Clifford unitaries to low gate counts is an active research area~\cite{Bravyi2021cliffordcircuit}. Importantly for their application to classical shadows}, since the $N$-qubit Clifford group constitutes a 3-design of the full unitary group~\cite{PhysRevA.96.062336}, this mapping can be analytically inverted by integrating over the full unitary group~\cite{Huang_classicalShadows}. This enables a reconstruction of the original density operator as
\begin{equation}
    \rho = \mathbb{E}_k[(2^N+1)U_k^\dagger |b_k \rangle \langle b_k|U_k - 1] . \label{eq:shadowstate}
\end{equation}

\toadd{The benefits of classical shadows have spurred its combination with a number of approaches in the field of electronic structure, sampling from the local Clifford \cite{Hadfield2022,PRXQuantum.6.010352}, matchgate \cite{PhysRevLett.127.110504,Wan2023_matchgate_shadows,zhao2025quantumclassicalauxiliaryfieldquantum} or global Clifford operator sets \cite{Huggins_unbiasingAFQMC,ren2025errormitigatednonorthogonalquantum,Blunt_fciqmc_w_shadows} to jointly evaluate expectation values. In comparison to other approaches using global Clifford shadows, in this work we aim to efficiently evaluate the energy (and other expectation values) of an eigenstate already prepared on a quantum device, in contrast to the use of the shadow state from a quantum device to facilitate a classical simulation, for example in defining trial states for suppressing the sign problem in (classically computed) Monte Carlo methods~\cite{Huggins_unbiasingAFQMC,Blunt_fciqmc_w_shadows} or to evaluate matrix elements for a classically tractable subspace expansion of the low-energy physics~\cite{ren2025errormitigatednonorthogonalquantum}.}

We can now define our operator set, $\{\hat{O}_j\}$, in order to extract the configurational amplitudes of interest, $c_{ij}^{ab}$ and $c_0$, from Eq.~\ref{eq:excitvec} in order to evaluate Eq.~\ref{eq:pert_energy}. These amplitudes are obtained as coherences of $\rho$, which we can extract akin to the approaches recently employed in QC-QMC approaches mentioned previously~\cite{Huggins_unbiasingAFQMC,Blunt_fciqmc_w_shadows}. To do this, rather than working with the density operator $\rho$, we construct $\tilde{\rho}=|\tau \rangle \langle \tau|$, defined by
\begin{equation}
    |\tau \rangle = \frac{1}{\sqrt{2}}(|0\rangle + |\Psi\rangle ),
\end{equation}
where $|0\rangle$ is the all-zero bit string, representing the (true) vacuum state of the system. If the state $|\Psi\rangle$ is generated by a circuit which preserves Hamming weight, as commonly performed in variational quantum eigensolver (VQE) algorithms~\cite{TILLY20221}, the generation of $|\tau\rangle$ can be achieved by initializing the circuit with an equal linear combination of $|0\rangle$ and another bit string with the desired Hamming weight~\cite{Blunt_fciqmc_w_shadows}. 
Since in Eq.~\ref{eq:Ham} we are working with electron number preserving Hamiltonians, the $|0\rangle$ state, representing a state with no electrons, is orthogonal to the target state $|\Psi\rangle$. We can then construct a set of $\mathcal{O}[N^4]$ Hermitian operators,
\begin{equation}
    \hat{O}_{ij}^{ab} = |0\rangle \langle \phi_{ij}^{ab}| + |\phi_{ij}^{ab}\rangle \langle 0 | , \label{eq:operators}
\end{equation}
where the states $|\phi_{ij}^{ab}\rangle$ correspond to single bitstrings in the canonical basis representing double excitations of the mean-field Slater determinant. Measuring these operators over the state $\tilde{\rho}$ allows for the extraction of the real part of the desired configurational amplitudes, as
\begin{equation}
    c_{ij}^{ab} = \langle \phi_{ij}^{ab}|\Psi \rangle = \mathrm{Tr}[\hat{O}_{ij}^{ab} \tilde{\rho}] , \label{eq:ampeval}
\end{equation}
while the imaginary part can also be extracted with a modification to these operators~\cite{Huggins_unbiasingAFQMC}.
Evaluating the terms in Eq.~\ref{eq:ampeval} with the classical shadow representation of the state $|\tau\rangle$ given in Eq.~\ref{eq:shadowstate}, we can efficiently construct the quantities of interest as the expectation value
\begin{equation}
    c_{ij}^{ab}=2(2^N+1)\mathbb{E}_k[\langle \phi_{ij}^{ab}|U_k^\dagger|b_k\rangle \langle b_k| U_k | 0 \rangle] . \label{eq:ampexpectation}
\end{equation}
Given the samples of the random Clifford operators, $U_k$, the matrix elements $\langle \phi_{ij}^{ab}|U_k^\dagger | b_k\rangle$ can be efficiently evaluated on classical computers by using the stabilizer formalism, requiring a classical time complexity of $\mathcal{O}[N^3]$; for further details see App.~\ref{app:stabilizer_calc}~\cite{PhysRevA.70.052328}. Furthermore, due to the non-Gaussian nature of the sampled distribution in Eq.~\ref{eq:shadowdensityoperator}, it has been shown that a `median-of-means' approach leads to a more effective bound on the expectation value than a simple mean~\cite{Huang_classicalShadows}. Therefore in this work we always consider the expectation value as a median, after dividing the samples into three groups and finding the mean of each group.

The classical shadow formulation therefore allows for the joint measurement of all $\mathcal{O}[N^4]$ operators given in Eq.~\ref{eq:operators}, from which the $c_{ij}^{ab}$ amplitudes required in Eq.~\ref{eq:pert_energy} can be computed, by using the same set of $M$ samples of $\{U_k, |b_k\rangle\}$. The precision of these resulting estimates depends on the `shadow norm' of the operators under the ensemble $\mathcal{U}=\mathcal{C}_N$ in Eq.~\ref{eq:shadowshots}, which can be considered as the maximum variance of the expectation values over the set of all possible states, denoted by $||\hat{O}_j||^2_\textrm{shadow}$. For the $\mathcal{C}_N$ ensemble, the shadow norm for the operators of interest in Eq.~\ref{eq:operators} is bounded by their Hilbert-Schmidt norm, $||\hat{O}_j||^2_\textrm{shadow} \leq 3\mathrm{Tr}[\hat{O}_j^2] = 6$. This norm is independent of $N$, and therefore we expect the precision by which we can extract any single $c_{ij}^{ab}$ amplitude to be independent of the system size. From Eq.~\ref{eq:shadowshots}, this results in the scaling in the number of shots for a maximum additive error in any amplitude over the entire set of required $c_{ij}^{ab}$ amplitudes ($\epsilon_\text{amp}$) as
\begin{equation}
    M \propto \frac{\log(N^4)}{\epsilon_\text{amp}^2} \propto\frac{\log(N)}{\epsilon_\text{amp}^2} .\label{eq:errorperamp}
\end{equation}
This implies a logarithmic scaling with system size in the number of shots required for an additive precision with which all configurational amplitudes can be estimated.

\subsection{Shot scaling of mixed energy estimator} \label{sec:shotscaling}

We can now consider the theoretical scaling of the number of shots required for a desired precision in the mixed energy estimate of Eq.~\ref{eq:pert_energy}, under the shadow sampling protocol and a mean-field trial state. If we assume that we can neglect covariances between the configurational amplitudes (found in Ref.~\cite{Kiser_2024} to be a good assumption for shadow sampled configurational amplitudes), and also assume a good trial state such that $c_0=\langle \Phi|\Psi\rangle \sim 1$, using Eq.~\ref{eq:errorperamp} we find the number of shots for an additive error, $\epsilon$, in the mixed energy estimate as
\begin{equation}
    M \propto \frac{\log(N)}{\epsilon^2} \sum_{ijab} |2v_{ijab}-v_{ijba}|^2 . \label{eq:shotsmixed}
\end{equation}
If we compare this to the shot number scaling for the direct energy estimate in Eq.~\ref{eq:shotspure}, we find a number of points of note:
\begin{itemize}
     \item The explicit scaling with system size, $N$, is substantially lower for the mixed estimator, scaling only logarithmically with system size compared to the $\mathcal{O}[N^4]$ in the direct sampling estimate of ungrouped Pauli operators of Eq.~\ref{eq:shotspure}. While this Pauli sampling scaling with system size can be reduced with various strategies \cite{C8SC05592K,jena2019paulipartitioningrespectgate,Crawford2021efficientquantum,PhysRevA.101.062322,9248636,doi:10.1021/acs.jctc.0c00008,10.1063/1.5141458,Wu2023overlappedgrouping,Yen2023,PRXQuantum.2.040320,PhysRevX.10.031064,Burns_2025}, it is always algebraic and scales at least as $\mathcal{O}[N^2]$ and often higher for {\em ab initio} systems in practice~\cite{PhysRevX.10.031064}. 
     \item The prefactor of the mixed estimator variance will generally be substantially smaller than the Pauli estimator, with the former depending only on the difference between electron repulsion terms in the Hamiltonian ($|v_{ijab}-v_{ijba}|^2$). This contrasts with the Pauli estimator variance, which encodes all the physics (including e.g. nuclear-electron repulsion and kinetic energy terms) of the total Hamiltonian in the $w_a$ weights of Eq.~\ref{eq:shotspure}. This is rationalized since only the {\em correlation} energy is estimated in the mixed estimator, rather than the {\em total} energy in the Pauli estimator. For most {\em ab initio} chemical and materials systems, the correlation energy makes up less than 1\% of the total energy, and therefore the mixed energy is a (generally far) smaller quantity, greatly accelerating the convergence with the number of shots. Empirically, we observe that both the mixed estimator variance prefactor of $\sum_{ijab}|2v_{ijab}-v_{ijba}|^2$ (Eq.~\ref{eq:shotsmixed}), and the Pauli estimator variance term of $\sum_a w_a^2$ (Eq.~\ref{eq:shotspure}), to scale approximately linearly with system size for Hydrogen chains under the Jordan-Wigner encoding. However, the mixed estimator term is substantially smaller and exhibits a reduced gradient in $N$, resulting in a reduction in the number of shots by a factor which also increases linearly with system size. For further details, see App.~\ref{app:ham_coeffs}.
     \item The effect of antisymmetry in the electronic state in the direct sampling of the energy is encoded in the Pauli strings, $\hat{P}_a$, of Eq.~\ref{eq:shotspure}. Depending on the fermion-to-qubit mapping employed, these strings will therefore in general contain a number of non-local and high-weight strings to encode this antisymmetry. In the mixed estimator however, this antisymmetry is encoded into the energy expression directly, via antisymmetric combinations of the Coulomb repulsion integrals (resulting in Coulomb- and exchange-like contributions), resulting in further reductions in the magnitude of the variance, and overall shot number~\cite{10.1063/1.4773819}. 
\end{itemize}

The analysis of the shot scaling for the mixed estimator is however likely too optimistic and simplistic. Firstly, there will be covariances between the terms of Eq.~\ref{eq:shotsmixed} which could result in an increased overall variance. Nevertheless, the covariances between configurational amplitudes sampled via classical shadows have been found to be small~\cite{Kiser_2024}. Covariances also occur with grouping strategies for joint measurements in direct sampling, so it may be fair to disregard their effect. Secondly, the mixed estimator of Eq.~\ref{eq:pert_energy} depends on the ratio of configurational amplitudes $c_{ij}^{ab}/c_0$, and we cannot necessarily assume that the magnitude of these ratios does not scale with system size. Covariances between the individual $c_{ij}^{ab}$ and $c_0$ estimates within a term could potentially lead to a bias in the overall energy in this ratio, though this was small enough to not be numerically observed in our experiments~\cite{PhysRevB.98.085118}. It also can be mitigated via a separate set of shadow samples for the $c_0$ amplitude or other approaches if required~\cite{Jackknife}. 
However, more importantly we expect that the magnitude of the configurational amplitudes will scale as $|\langle \phi | \Psi \rangle| \sim 2^{-N}$ for a strongly correlated state which lacks compact support on the configurational space, giving rise to an exponentially decreasing magnitude with system size. 

Since we are interested in a fixed additive error in the ratio of two amplitudes which are decreasing in size, it is more appropriate to require a fixed {\em relative} precision of the individual amplitudes. This introduces a factor which depends on the overlap with the trial state into the estimate of the number of shots in Eq.~\ref{eq:shotsmixed}, and hence the additive error in the mixed energy estimator now depends on $|\langle \Phi | \Psi \rangle|^{-2}$; see App.~\ref{app:full_error} for more details.
Formally, the overlap of any approximate trial state with a true eigenstate will decrease exponentially in the large system limit, resulting in an implicit exponential scaling in the number of shots to evaluate the mixed estimator with system size~\cite{simon2025dividingconqueringvanvleck}. This changes our scaling in shot number from $\mathcal{O}[N\log(N)]$ to exponential in system size. This argument has also previously been leveled at other approaches where classical algorithms have been combined with quantum-derived information, most notably the QC-QMC methods~\cite{Huggins_unbiasingAFQMC}, where this issue of sampling ratios of exponentially vanishing amplitudes also appears~\cite{mazzola2022exponentialchallengesunbiasingquantum}.

It is clear therefore that Eq.~\ref{eq:shotsmixed} does not provide the full picture, and some factor which depends on the overlap with the trial state is also expected. While the exponential increase in variance with system size therefore follows as a formal asymptotic argument, these arguments can only take us so far in an understanding of whether the approach can provide practical benefits and algorithmic quantum advantage~\cite{zimboras2025mythsquantumcomputationfault,lee2022responseexponentialchallengesunbiasing}. Eigenstate simulation is in the Quantum Merlin-Arthur (QMA) complexity class (loosely, the quantum analogue to the NP class) precisely because of the presence of an exponentially decaying overlap with an initial state in the preparation phase (which results in the same formal scaling for QPE with system size as $|\langle \Phi | \Psi \rangle|^2$)~\cite{Lee2023}. Nevertheless, we can still hope to find improvements in algorithms before this asymptotic limit, and we must analyze the performance numerically to understand the scope of the approach.

Furthermore, we can extend this approach to optimize the mean-field state to maximize overlap, or consider more accurate trial states which can still be efficiently encoded on a quantum device, with a natural choice being tensor network states~\cite{berry2024rapidinitialstatepreparation,doi:10.1021/acscentsci.8b00788,zou2025multireferenceerrormitigationquantum}. This is an area of active research, to extend the scope of state preparation protocols on quantum computers for eigenstate preparation and beyond, and these developments are likely to be transferrable to this setting. Nevertheless, we will continue with a simple Hartree--Fock state here, which has already proven effective in state preparation for relatively simple chemical systems~\cite{tubman2018postponingorthogonalitycatastropheefficient,PRXQuantum.3.010318}, and numerically analyze the efficiency of the mixed estimator compared to direct sampling of the Hamiltonian as the overlap with this single determinant trial state changes.

\subsection{Numerical performance of mixed estimator}

\begin{figure*}[htb!]
    \centering
    \begin{subfigure}[t]{0.5\textwidth}
        \centering
        \includegraphics[height=6.5cm]{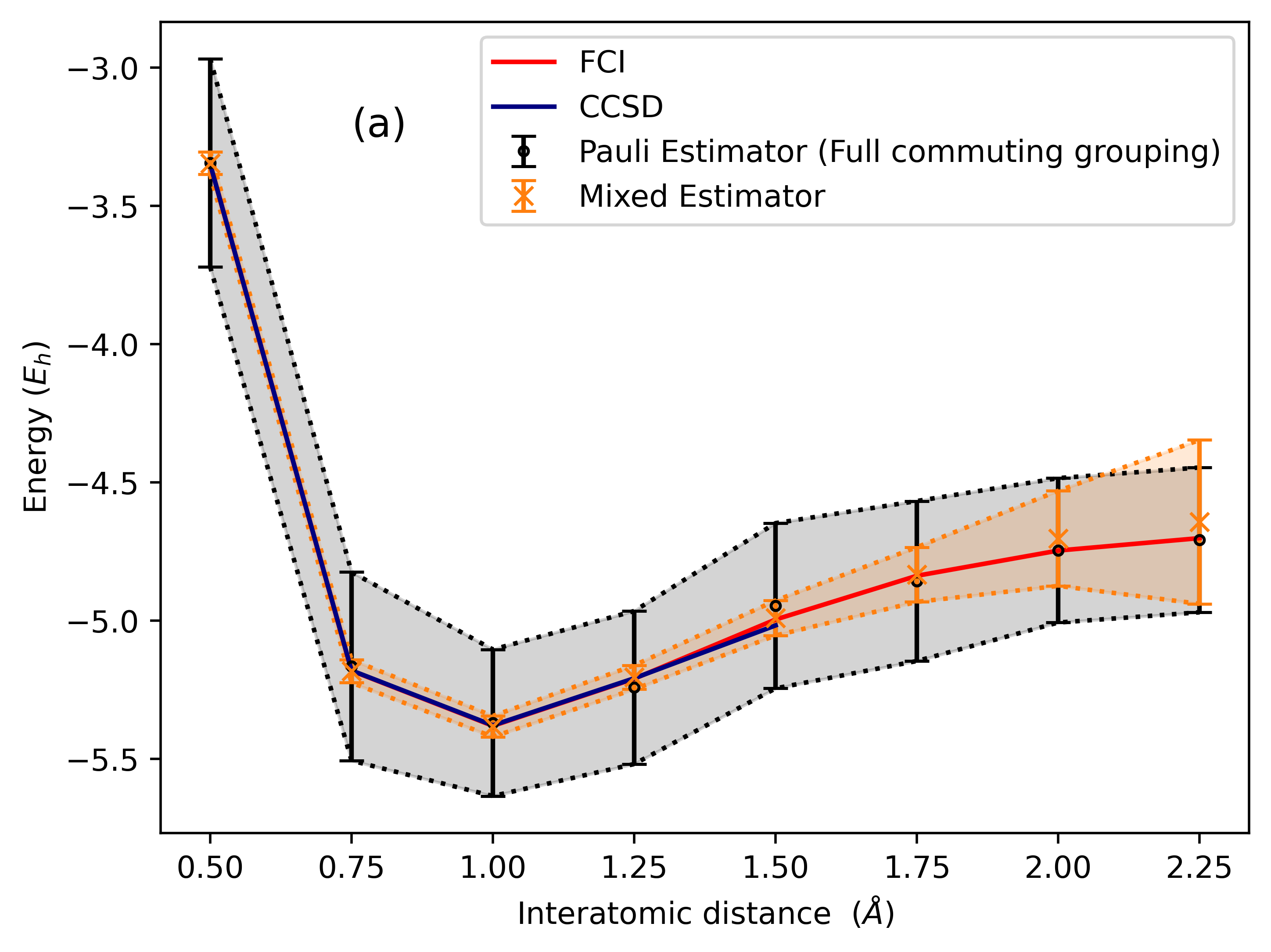}
    \end{subfigure}%
    ~
    \begin{subfigure}[t]{0.5\textwidth}
        \centering
        \includegraphics[height=6.5cm]{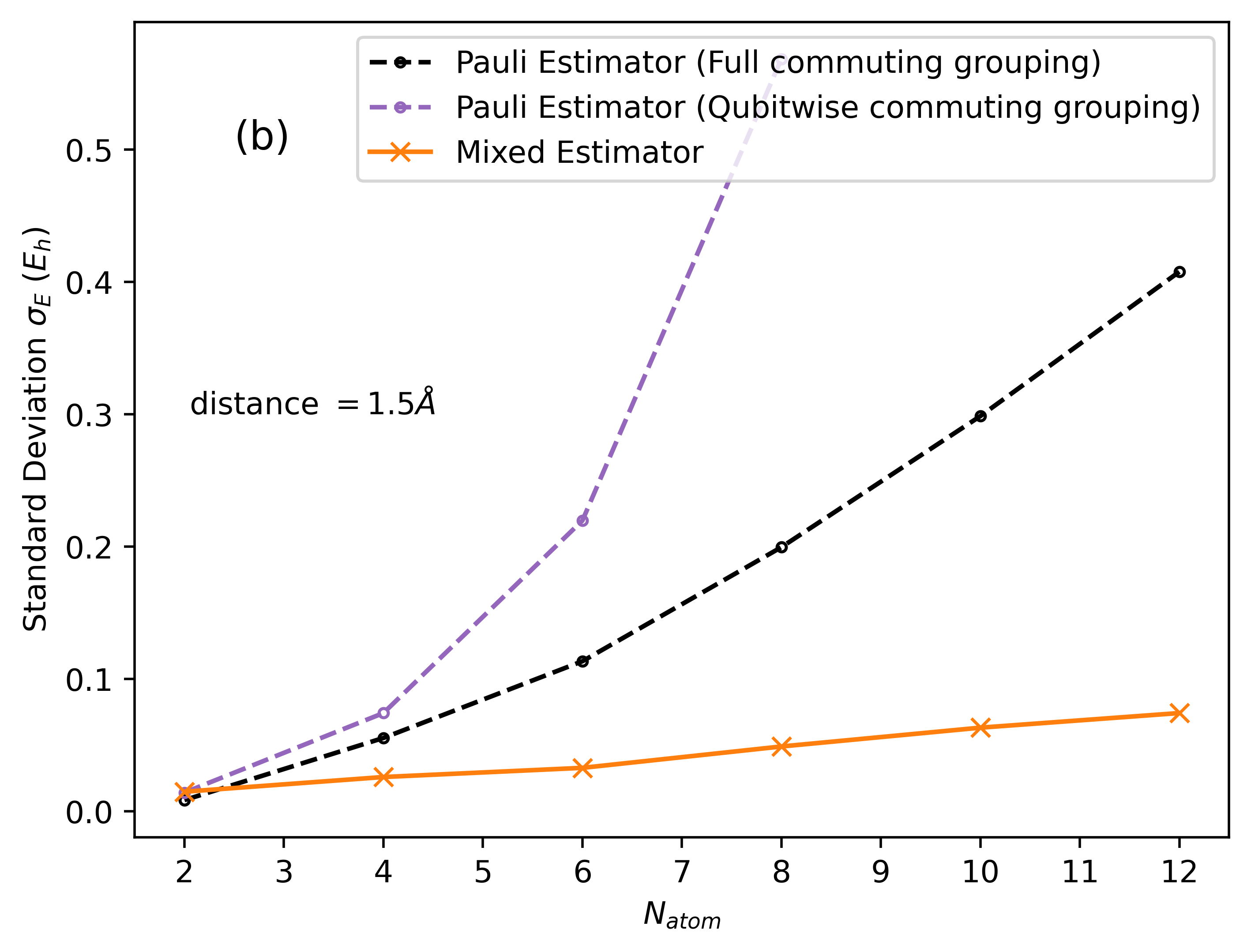}
    \end{subfigure}%
    \caption{{\bf Variance of mixed and Pauli energy estimators for Hydrogen chains over changing correlation strengths and system size.} (a) Mean and standard deviation of the energy estimate distributions of the Pauli and mixed estimators sampled with 1000 shots, for a ten atom symmetrically stretched Hydrogen system in an STO-3G basis. Also included is the exact solution (FCI, red) and coupled-cluster with singles and doubles (CCSD, navy) results, which failed to converge for distances above $1.5$\AA. Both the Pauli and mixed estimators are obtained from samples of the exact ground statevector, with the Pauli estimate fully grouping the Pauli operators into commuting sets using a graph coloring algorithm, and the mixed estimate employing the Clifford shadow protocol as described in Sec.~\ref{sec:shadowsampling}. (b) Standard deviations of the Pauli (dashed) and mixed (solid) energy estimator distributions as the Hydrogen chain system size increases for a fixed interatomic distance of $1.5$\AA, and a fixed 1000 shots per sample of the distribution. The black line again shows the Pauli operators fully grouped into as few commuting sets as possible, while the purple results group Pauli operators based on qubitwise commutivity. The orange line represents the standard deviations of the mixed estimator with the Hartree--Fock trial state using the Clifford shadows method, and is shown to exhibit approximately linear scaling of the error with system size.}
    \label{fig:hstretch_errors}
\end{figure*}


We compare the energy estimators applied to the symmetric stretching coordinate of a linear chain of Hydrogen atoms, without breaking any spatial or spin symmetries (i.e. a restricted basis). This is an important and widespread benchmark model for algorithmic development in {\em ab initio} systems. It features the full complexity of a realistic Hamiltonian with a correlation strength which can be tuned from weakly correlated at compressed geometries to strongly correlated at stretched cases~\cite{Motta_hydrogen, Huggins2021}. Figure \ref{fig:hstretch_errors}(a) shows the potential energy over the stretching of the chain for $10$ atoms (20 qubits) in a minimal basis, comparing the energy and its standard deviation for the mixed and Pauli estimators while the number of shots is fixed at $1000$ for both methods. In addition, two reference classical methods are shown, with full configuration interaction (FCI) giving the exact solution, and coupled-cluster with single and double excitations (CCSD), providing an inexact yet classically tractable ansatz for comparison. Since CCSD is a resummed perturbative expansion, it fails for strongly correlated systems. Therefore, it provides excellent agreement with FCI for compressed geometries, but fails to converge on stretching beyond a bond length of $1.5$\AA, which can be considered as the point at which strong correlations lead to the breakdown of standard perturbative approaches.


The Pauli and mixed energy estimates are sampled from the exact ground state via statevector simulation. \toadd{This ensures that there is no bias in either the mixed estimator or Pauli estimator results, and therefore we can compare the variance of the estimators in isolation.}
The Pauli estimator is given as the mean and standard deviation of the distribution of 1000 shot estimates (with the estimated standard deviation taken as the average of 100 simulations to lower the uncertainty in the true distribution). Commuting Pauli strings from the Jordan-Wigner encoding of the Hamiltonian are fully grouped using a heuristic graph coloring algorithm to find the optimal cliques~\cite{10.1063/1.5141458}. This represents the standard approach to sampling the energy, with the version of this procedure restricted to qubitwise commuting groups the default in Qiskit Aer 0.16.1. The mixed estimator results are found with the same number of shots, but via the classical shadow procedure of Sec.~\ref{sec:shadowsampling}. We note that it is also possible to directly sample the expectation values of the operators of Eq.~\ref{eq:operators} required for the mixed estimator thereby avoiding the requirement of the classical shadow protocol entirely. Results from this are shown in App.~\ref{app:directconfigsampling}, and demonstrate the advantage which comes from the combination of both the classical shadow protocol with the mixed estimator.


While both sampled energy expectation values in Fig.~\ref{fig:hstretch_errors}(a) are numerically unbiased compared to the FCI reference, the results show that the Pauli estimator has an approximately constant standard deviation over the different bond lengths. This reflects that the variance in Eq.~\ref{eq:shotspure} is not explicitly dependent on the strength of the correlations in the system, although there can be indirect effects through the variance of the individual Pauli operators. This is in contrast to the mixed estimator, where the estimate in Eq.~\ref{eq:mixedestimator} has an explicit dependence on the overlap with the trial state $\ket{\phi}$, which here is the Hartree--Fock state. As a result, as the chain is stretched, the variance of the energy estimate increases, as strong correlations reduce the overlap with the trial state to small values. 
Nevertheless, it is only at the final point in the near-dissociated atomic limit where we find the mixed estimator to exhibit a larger variance in the energy estimate than the Pauli estimator. Around equilibrium geometries of $1.0$\AA, the standard deviation of the energy distribution with 1000 shots is nearly an order of magnitude smaller. Due to the $\epsilon^{-2}$ scaling in Eqs.~\ref{eq:shotspure} and \ref{eq:shotsmixed}, this leads to a more than 50-fold reduction in the number of shots required to extract the energy with a desired error when using the mixed compared to the Pauli estimator here. The reduction in the statistical error for the mixed estimator persists well into the regime where the chain is stretched beyond equilibrium, including the non-perturbative regime beyond $1.5$\AA~ where CCSD fails to converge. The estimators exhibit similar errors only in highly correlated settings once the overlap with the trial state has dropped below $0.25$. 

We can also consider the scaling of the error with system size. The naive arguments of Sec.~\ref{sec:shotscaling} suggest a scaling in the number of shots as a function of system size for a constant error is $\mathcal{O}[N\log(N)]$ for the mixed estimator, while up to potentially $\mathcal{O}[N^5]$ for the Pauli estimator (with the extra factor of $N$ arising from the empirical scaling of the Hamiltonian-dependent prefactors shown in App.~\ref{app:ham_coeffs}). However, this ignores the implicit exponential scaling of the mixed estimator due to the decreasing overlap with the trial state as the system grows, as discussed above. 
Figure~\ref{fig:hstretch_errors}(b) therefore shows the standard deviation of the two energy estimates for samples of $1000$ shots while increasing the number of atoms in the chain. The interatomic spacing is fixed at $1.5$\AA~ for all chain lengths -- the point at which CCSD starts to fail for the ten atom system, and with the weight of the Hartree--Fock configuration $\approx0.6$ for 12 atoms, indicating a moderately strongly correlated state. 

The dashed lines of Fig.~\ref{fig:hstretch_errors}(b) show the standard deviation of the Pauli estimator with two different graph-coloring based approaches to group mutually commuting Pauli strings, either attempting to fully group all terms (black) or restricting to qubitwise commutation (purple). Exploiting qubitwise commutativity only reduces the prefactor in the number of groups of Pauli strings to measure and therefore still retains $\mathcal{O}[N^4]$ terms in keeping with the original Hamiltonian~\cite{10.1063/1.5141458}, while a full grouping strategy is observed to lead to a $\mathcal{O}[N^3]$ scaling in the number of groups to measure~\cite{9248636}. Both clearly exhibit a beyond-linear polynomial scaling in system size, commensurate with these scalings. In contrast, the standard deviation of the mixed estimator exhibits an approximately linear increasing standard deviation in the estimate for a fixed shot budget. It is clear that for this range of system sizes, the naive scaling arguments hold, and that a substantial reduction in the number of samples is possible for a given energy error. The mixed estimator is therefore likely to be effective for system sizes and correlation strengths beyond classically tractable approaches.

\subsection{Noisy and approximate states}

\begin{figure*}[htb!]
\begin{center}
    \centering
        \begin{subfigure}[t]{0.5\textwidth}
            \centering
            \includegraphics[height=6.5cm]{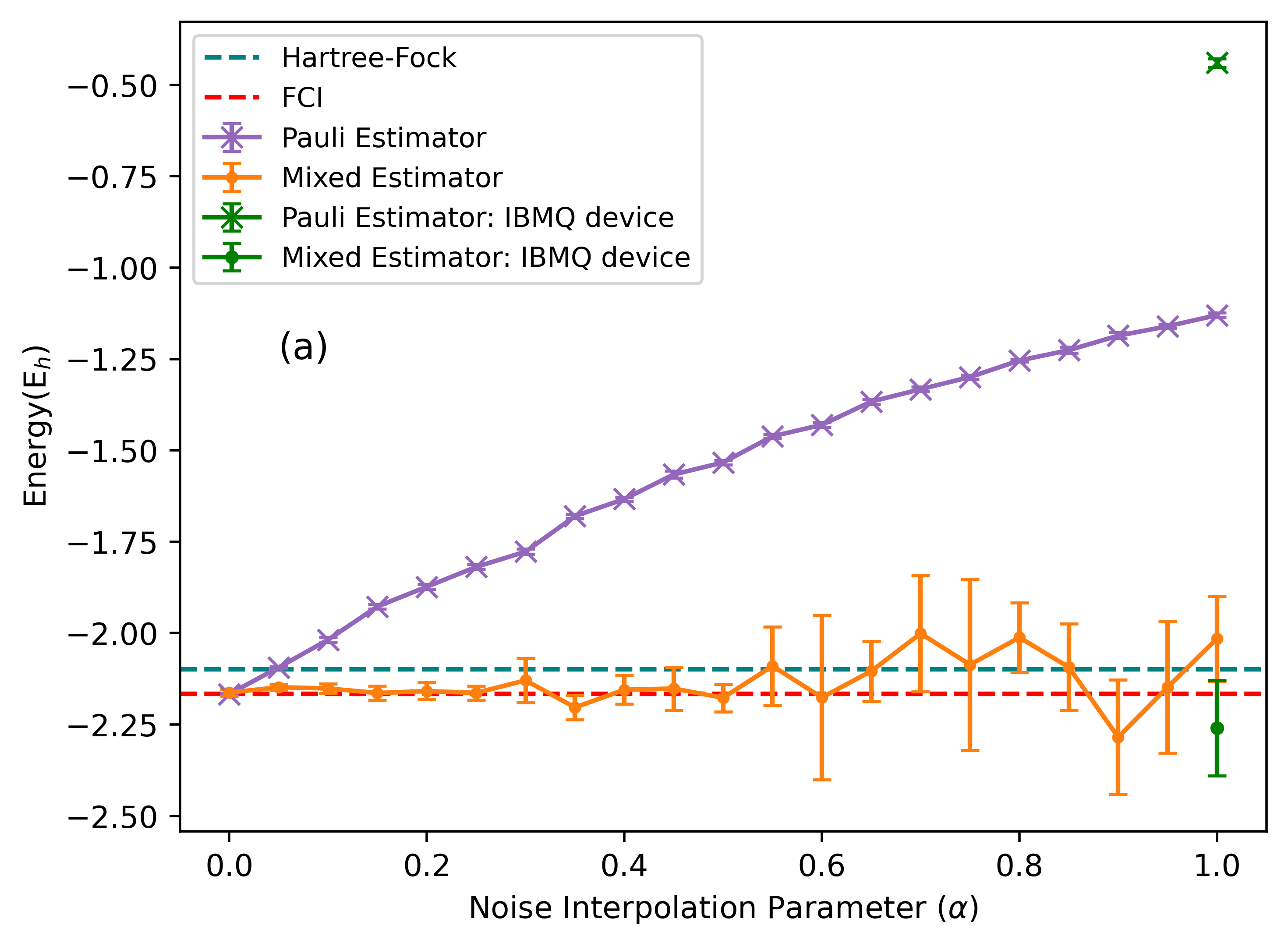}
        \end{subfigure}%
        ~
        \begin{subfigure}[t]{0.5\textwidth}
            \centering
            \includegraphics[height=6.5cm]{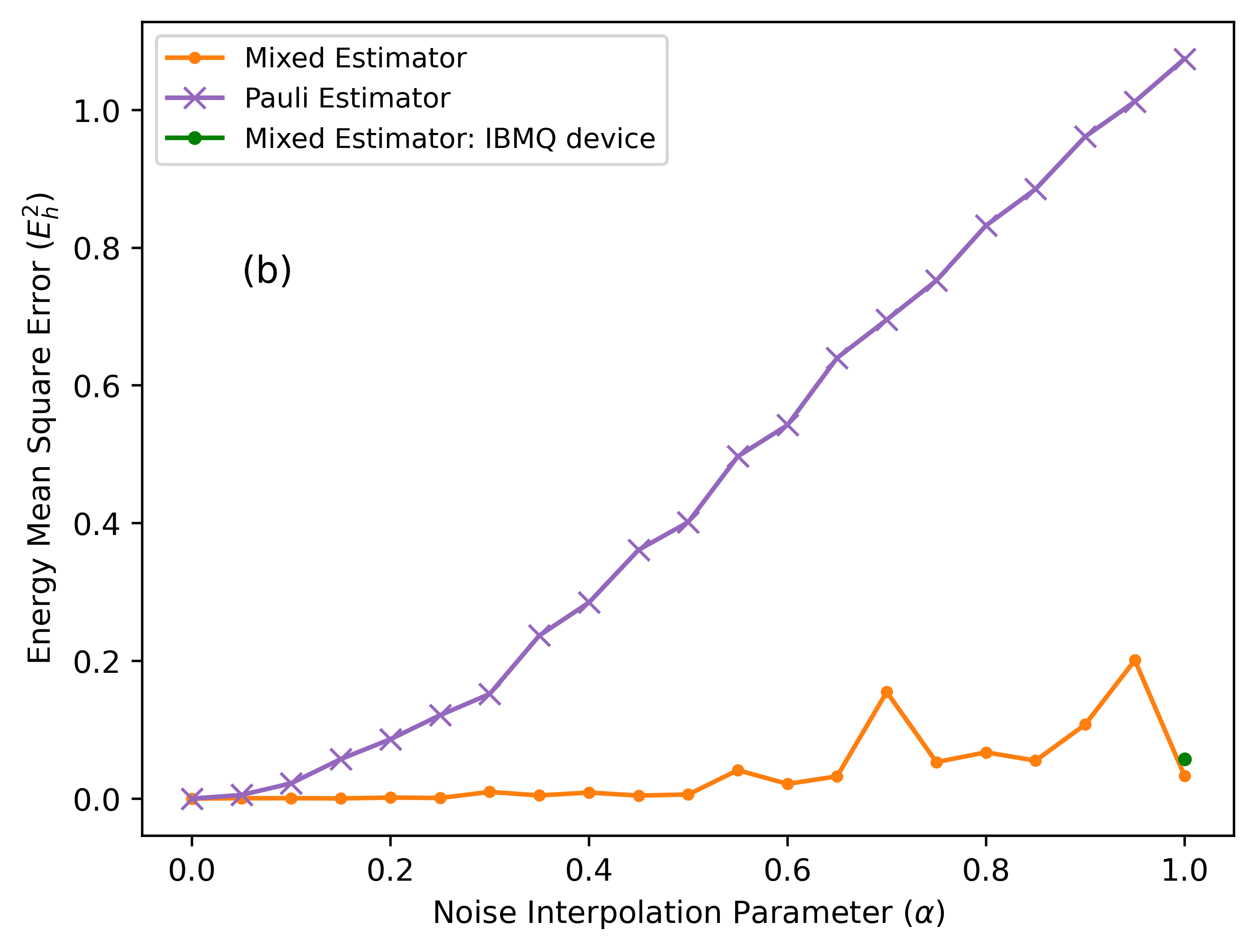}
        \end{subfigure}%
    \caption{{\bf The effect of noise on the energy estimators for a fixed, optimized VQE ansatz, showing improved resilience of the mixed estimator.} (a) Energies for a four atom Hydrogen chain, computed via the Pauli (purple) and mixed (orange) estimators for different noise levels, each using eight sets of 1000 shots. \toadd{The points are the mean energy over these sets, characterizing the bias in the estimators due to the noise and inexact state, while the error bars characterize the standard error over these sets for the different estimators, indicating the statistical variance of the results due to finite shot numbers.} The green points represent execution on the IBM Brisbane quantum device with the same number of shots, \toadd{somewhat} validating the performance of the simulated results at $\alpha=1$. The exact energy (FCI, dashed red) and Hartree--Fock energy (dashed light blue) are shown for comparison. The noise model for a given $\alpha$ is shown in Eq.~\ref{eq:noise_channel_interpolation}, with $\alpha=0$ a noiseless simulation, and $\alpha=1$ approximating the noise of the real quantum device. \toadd{(b) The mean-squared error (MSE) for the different estimators compared to the exact energy, incorporating both the bias and variance of the estimators in a single metric.}}
    \label{fig:noise}
\end{center}
\end{figure*}

To gauge the applicability to realistic quantum devices, we consider the resilience of the mixed energy estimate with respect to noise and sampling of inexact states. 
While Eq.~\ref{eq:mixedestimator} must agree with the exact energy when the state $\ket{\Psi}$ is an eigenstate of the Hamiltonian, it will not in general agree with the Pauli estimate of Eq.~\ref{eq:pureestimate} for other states\toadd{, and therefore we need to consider both the bias and variance of the estimators in the presence of inexact and noisy states}.
We consider the ground state of a four atom (8 qubit) Hydrogen chain obtained from the Variational Quantum Eigensolver (VQE), using the Local Unitary Cluster Jastrow (LUCJ) ansatz \cite{LUCJ_ansatz} with $23$ parameters, implemented and optimized using the {\tt ffsim} package \cite{ffsim}. The restriction of the state to the employed ansatz leads to an inexact state with an energy error of $\sim5\textrm{mE}_h$ after optimization. 

The optimized circuit was compiled for the IBMQ architecture 
with a custom noise model \toadd{constructed to best reproduce the publicly available physical device error metrics. These included using the $T_1$ and $T_2$ relaxation times to parameterize a single-qubit thermal relaxation channel introducing phase and amplitude damping, one- and two-qubit gate error rates used to inject local depolarization, and local read out error rates designed to match the the IBM Eagle device, giving an overall noise channel, $\mathcal{M}$.} This error model was then linearly interpolated via a single parameter, $\alpha$, to the identity channel ($\mathcal{I}$) as
\begin{equation}\label{eq:noise_channel_interpolation}
    \mathcal{M}'(\alpha) = \mathcal{I} - \alpha (\mathcal{I} - \mathcal{M}) ,
\end{equation}
resulting in the noiseless identity channel when $\alpha=0$ (representing ideal, error-free quantum operations), and the physically modeled noise channel when $\alpha=1$ (representing the noise model of the IBMQ Eagle architecture).

In Fig.~\ref{fig:noise} we plot the energy of this system estimated from eight sets of 1000 shots, as the noise model is varied between $0 \leq \alpha \leq 1$, to determine the mean and the error bar over the groups. We plot the Hartree-Fock energy (light blue) and exact FCI energy (red) for comparison and the results from the mixed estimator (orange) and the default IBM Pauli estimator for device execution (purple). 
We also validate the results for the full noise model ($\alpha=1$) by using the physical IBM Brisbane quantum device (green), which has the same processor architecture as the Eagle device used for the noise model. All measurements are taken from the same fixed VQE circuit (optimized via statevector simulation in the absence of noise) defining the state of interest. The physical quantum hardware results qualitatively agree with our noisy statevector simulations, validating the noise model employed. \toadd{In the right hand plot, we also show the mean squared energy error (MSE) compared to the exact results, combining the effects of both the bias and the variance in the estimators for this inexact energy, and demonstrating the combined benefit of the mixed estimator.}

We find that as $\alpha$ increases from the noiseless simulation, the \toadd{bias} in the Pauli estimator increases rapidly, such that the standard approach to estimating the energy as $\mathrm{Tr(\rho \mathcal{H})}$ is worse than the Hartree--Fock energy, even with $\alpha$ as low as $\sim0.05$. As the noise continues to grow, the estimate becomes significantly worse than even a mean-field description of the electronic structure. This is due to an accumulation of errors in the circuit, despite the deliberately shallow LUCJ state ansatz, with all errors necessarily raising the energy variationally as the error rate increases. 
In contrast, the mixed estimator maintains a \toadd{less biased estimate, with the} average much closer to the true value of the noiseless state, with increasing noise levels reflected primarily in a larger variance in the estimate, with a degree of error cancellation afforded by the non-variationality of the approach. In the large noise limit, the state will have a random overlap with the trial state in Eq.~\ref{eq:mixedexplicit}, providing an estimate given by the trial state energy (Hartree--Fock) plus some random change whose variance grows as the overlap decreases due to the noise. However, this effectively bounds the systematic error by the Hartree--Fock energy, and it is clear that there is a range of noise with $\alpha \lesssim 0.5$ where the mixed energy result remains a good approximation to the FCI result, and provides a statistically significant portion of the correlation energy commensurate with the LUCJ description. This is despite the formal Pauli estimator providing an energy substantially worse than the mean-field result, and therefore negative correlation energies.

It has been noted in other contexts that estimates of these ratios of overlaps required for the mixed estimator benefit from an inherent resilience to noise, both in the stochastic shadow sampling procedure \cite{PhysRevResearch.6.043063}, and to global depolarizing noise \cite{Huggins_unbiasingAFQMC,Blunt_fciqmc_w_shadows}, which can often be considered a reasonable approximation to the cumulative effect of local noise \cite{Arute2019,dalzell2021randomquantumcircuitstransform}. \toadd{This is because global depolarizing noise will exactly cancel in the ratio of probability amplitudes \cite{Huggins_unbiasingAFQMC}, and therefore leave the mixed estimator unchanged. While our simulated noise model goes well beyond just global depolarization,} this inherent resilience in the mixed estimator means that a reduction in the noise of the quantum device by as little as $2-3$ times could allow a reasonable energy estimate to be obtained without error mitigation in many cases. This provides the possibility of extracting faithful correlation energies in noise regimes an order of magnitude larger than those required for meaningful Pauli energy estimates. While this partially obviates the need for error mitigation methods, their combination within the mixed estimator to further improve noise resilience remains an interesting route for further development~\cite{zou2025multireferenceerrormitigationquantum}. However, as noted before, a drawback of the mixed energy functional is that it is unbounded, and therefore cannot be used to directly optimize a variational ansatz e.g. within VQE. Nevertheless, the significantly enhanced robustness to noise will still likely make it an attractive proposition in estimating the energy of already obtained states, both in the final step of a VQE workflow, and within other quantum algorithms.

\section{Non-energetic properties} \label{sec:properties}
\begin{figure*}[ht!]
    \centering
    \begin{subfigure}[t]{0.5\textwidth}
        \centering
        \includegraphics[height=6.5cm]{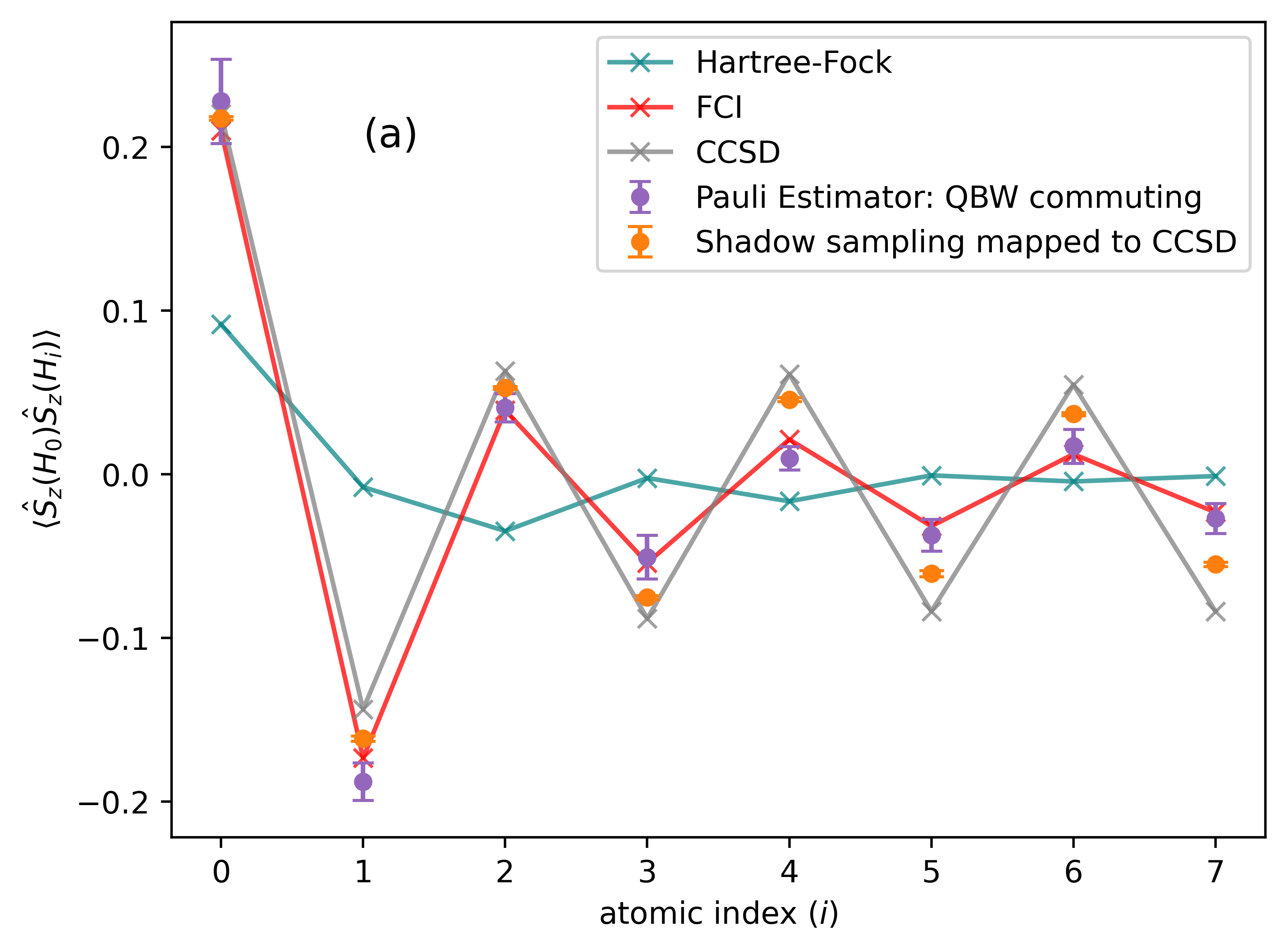}
    \end{subfigure}%
    ~
    \begin{subfigure}[t]{0.5\textwidth}
        \centering
        \includegraphics[height=6.5cm]{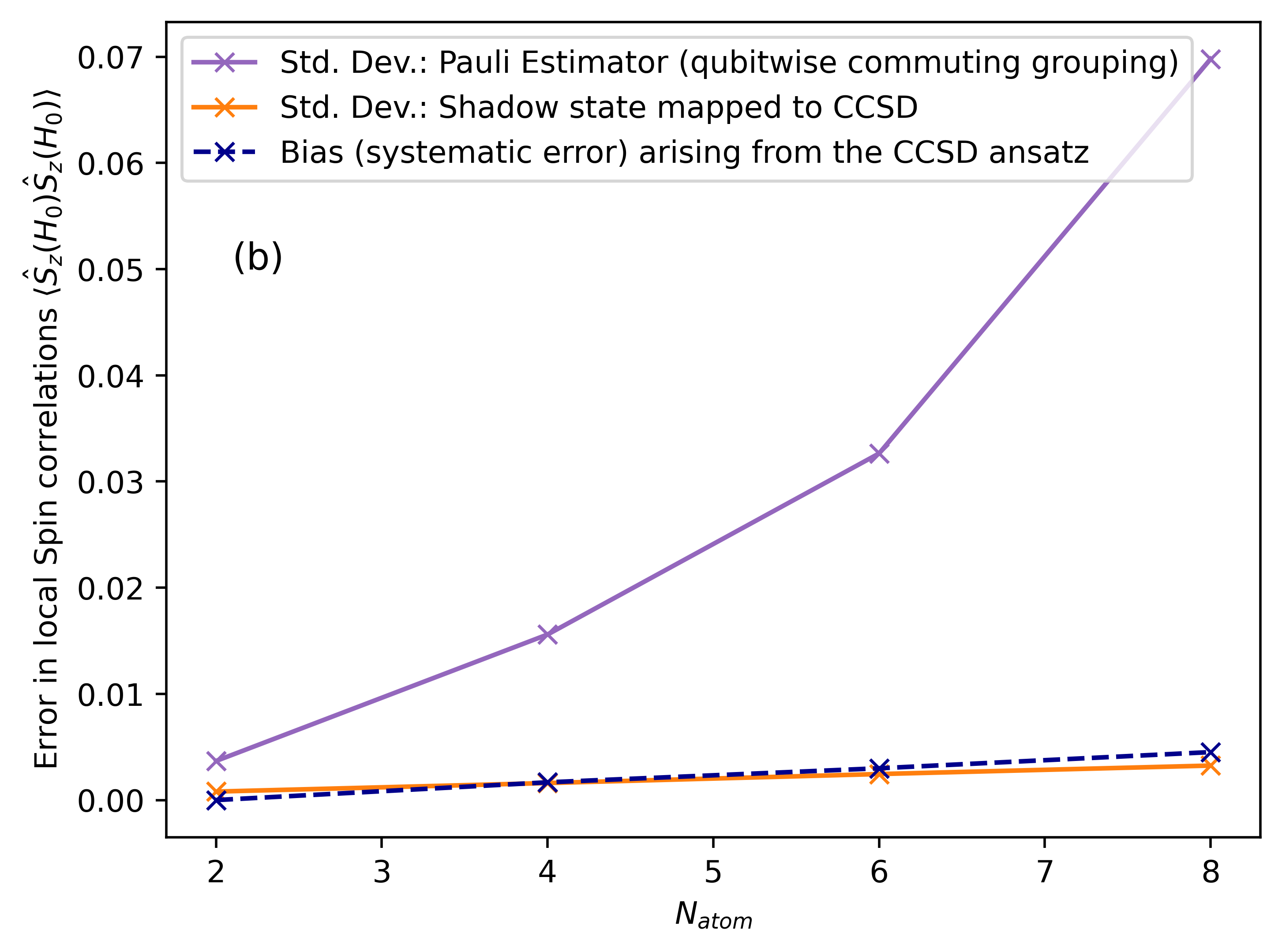}
    \end{subfigure}%
    \caption{{\bf Systematic and random errors in the magnetic correlations of Hydrogen chains} (a) The spin-spin correlation function of an eight atom hydrogen chain \toadd{(symmetrically stretched to $1.5$\AA)} between the first atom and its $i^\text{th}$ nearest neighbor in a STO-3G basis. Hartree--Fock (light blue), FCI (red), and CCSD (grey) are shown as classical methods, while ten sets of 40,960 shots are used to estimate the mean and error of the expectation values from the exact statevector, comparing a direct sampling of qubitwise grouped Pauli strings (purple) and a mapping of the shadow state to a CCSD ansatz (orange). (b) The local (same-atom) spin-spin correlation function for an N$_{\mathrm{atom}}$ hydrogen chain, showing the standard deviation of the distribution of expectation values with 40,960 shots for the two approaches, with the navy dashed line also showing the systematic error in the CCSD model for this property when the exact coefficients are used, quantifying the \toadd{bias} introduced by this CCSD model parameterization.}
    \label{fig:SzSz}
\end{figure*}

While the mixed estimator offers an attractive alternative to Pauli sampling, in the form presented, it is only suitable for energetic expectation values (or operators that commute with the Hamiltonian). We now consider a natural and important extension as to whether other properties of the state can be extracted using the {\em same} configurational amplitude information that was sampled for the mixed energy estimate, without requiring further measurements on the quantum device. This is important for electronic structure simulation, where properties such as electrical moments, magnetic order and correlation functions are often as important as energetics for understanding material behavior. To achieve this, we map the excitation amplitudes of Eq.~\ref{eq:excitvec} extracted from the shadow state to a coupled cluster ansatz. The coupled cluster ansatz takes the form
\begin{equation}
    |\Psi_{\text{cc}} \rangle = e^{\hat{T}}|\phi \rangle , \label{eq:cc}
\end{equation}
where $|\phi\rangle$ is a single determinant state, and $\hat{T}$ is an operator encoding the amplitudes of the state. Here we truncate this to single and double excitations, to define the CCSD ansatz~\cite{RevModPhys.79.291}, as
\begin{equation}
    {\hat T} = \sum_{ia} t_i^a {\hat c}^{\dagger}_a {\hat c}_i + \sum_{ijab} t_{ij}^{ab} \hat{c}^{\dagger}_a \hat{c}^{\dagger}_b \hat{c}_i \hat{c}_j . \label{eq:tamps}
\end{equation}
The previously obtained single and double excitation amplitudes, $c_i^a$ and $c_{ij}^{ab}$, extracted from our quantum state, $|\Psi\rangle$, can be mapped to the amplitudes of $t_i^a$ and $t_{ij}^{ab}$ required to define the state $|\Psi_{\text{cc}}\rangle$ in a closed form procedure on classical resources~\cite{RevModPhys.79.291}. In particular, these $t$-amplitudes can be constructed such that the projection of the sampled shadow state onto the single and double excitations of the reference single-determinant state, $|\phi\rangle$, are equal to the same projections from the coupled cluster model state, i.e. $t_i^a$ and $t_{ij}^{ab}$ are defined such that $\langle \phi_{ij}^{ab}|\Psi \rangle=\langle \phi_{ij}^{ab}|\Psi_\text{cc}\rangle$ (and similarly for the single excitations). However, the exponential form of the coupled cluster ansatz in Eq.~\ref{eq:cc} means that the $|\Psi_\text{cc}\rangle$ state found in this way will have probability amplitudes over all configurations in the Hilbert space, derived as disconnected products of the single and double excitation amplitudes. 

This allows us to use the previously obtained excitation amplitudes from the shadow state in order to define this classically tractable coupled cluster surrogate model, from which properties of interest can be computed efficiently. Importantly, this coupled cluster $|\Psi_\text{cc}\rangle$ state will share the same energy as the state $|\Psi \rangle$, which will therefore be exact if the exact excitation amplitudes $c_i^a$ and $c_{ij}^{ab}$ are used for this mapping. This property is exploited in `tailored' coupled cluster approaches, which have recently been combined with quantum algorithms in order to introduce quantum-derived information within a low-energy subspace into a classical CCSD state~\cite{doi:10.1021/acs.jctc.4c00037,PhysRevResearch.6.023230}. However, in this work we are defining a coupled cluster state parameterization over the entire space as a surrogate model of the quantum state, and considering the accuracy to which properties can be estimated from this \toadd{classically} tractable form. 

In particular, we can focus on the two-body reduced density matrix (2-RDM), from which all (static) low-body expectation values of interest can be derived. As such, their efficient extraction from quantum states is an active research endeavor, motivating the development of matchgate shadow tomography and other approaches~\cite{PhysRevLett.127.110504,Crawford2021efficientquantum,Bonet-Monroig_majorana_clique,Hamamura2020,PhysRevResearch.3.033230}. The 2-RDM is defined as
\begin{equation}
    \Gamma_{\alpha\beta\gamma\delta} = \braket{\Psi| \hat{c}^{\dagger}_{\alpha} \hat{c}^{\dagger}_{\beta} \hat{c}_{\delta} \hat{c}_{\gamma}|\Psi}.
\end{equation}
Given a definition of the $\hat{T}$ operator of $|\Psi_\text{cc}\rangle$ mapped from the shadow state, the 2-RDM of this surrogate model can be computed efficiently on classical resources as
\begin{equation}
    \Gamma_{\alpha\beta\gamma\delta} = \braket{\phi|(1+\hat{\Lambda})e^{-\hat{T}} \hat{c}^{\dagger}_{\alpha} \hat{c}^{\dagger}_{\beta} \hat{c}_{\delta} \hat{c}_{\gamma} e^{\hat{T}} | \phi} , \label{eq:ccrdm}
\end{equation}
where $\hat{\Lambda}$ are a set of de-excitation operators, defined as the hermitian conjugate of the excitation operators in Eq.~\ref{eq:tamps}. These de-excitation amplitudes can be classically optimized in the presence of the fixed $\hat{T}$ operator in the expression above, as Lagrange multipliers, in order to ensure a correct bivariational form for the theory~\cite{RevModPhys.79.291}. The full procedure therefore involves first extracting the excitation amplitudes $c_i^a$ and $c_{ij}^{ab}$ from the state on the quantum device via shadow tomography, mapping them to define the $t$-amplitudes of Eq.~\ref{eq:tamps}, before optimizing the $\hat{\Lambda}$ de-excitation amplitudes and evaluating the 2-RDM of Eq.~\ref{eq:ccrdm} on a classical device, for which we use the {\tt ebcc} package~\cite{Backhouse_ebcc_Coupled_cluster_2024}.

Due to the truncation of the cluster operator to only single and double excitations in this work (Eq.~\ref{eq:tamps}), the properties derived from this classical surrogate state will incur a systematic error from the incomplete state definition, even if the excitation amplitudes from which the state is constructed are themselves exact. This contrasts with the properties of the mixed energy estimator. Nevertheless, the accuracy by which the properties can be extracted in principle surpasses that of a classical optimization of the coupled-cluster state, since the amplitudes are appropriately relaxed due to the correlations which manifest beyond the truncated excitation level considered. Furthermore, it is possible to systematically alleviate the constraints of the model by increasing rank of excitations which are considered in the $\hat{T}$ operator of Eq.~\ref{eq:tamps}, which would become \toadd{unbiased} in the case that all excitations are considered. However, we restrict to a single and double excitation parameterization of the coupled-cluster model in this work.

To numerically test the performance of these non-energetic properties from this CCSD surrogate model, we return to the chain of eight Hydrogen atoms and consider the magnetic spin correlation functions which emerge from the correlated electronic structure upon symmetrically stretching the atoms to $1.5$\AA. This question has been heavily studied by classical electronic structure methods to characterize long-range magnetic order in this system~\cite{Motta_hydrogen}, and is a central observable in many (e.g. neutron scattering) experimental setups. We define the two-point spin correlation function between two Hydrogen atoms, $H_i$ and $H_j$, as $\braket{\hat{S}_z(H_i) \hat{S}_z(H_j)}$, with the spin magnetization operator for the atom $H_i$ defined as
\begin{equation}
    \hat{S}_z(H_i) = \frac{1}{2}(\hat{c}^\dagger_{i,\uparrow} \hat{c}_{i,\uparrow} - \hat{c}^\dagger_{i,\downarrow} \hat{c}_{i,\downarrow}),
\end{equation}
where $i$ labels the (L\"owdin orthonormalized) local atomic orbital of the atom $H_i$. This correlation function will describe the existence of any (e.g. antiferromagnetic) spin ordering that spontaneously appears in the system due to the correlated quantum fluctuations.

In Fig.~\ref{fig:SzSz}(a) we calculate this spin-spin correlation function between the first atom and increasingly distant atoms in the chain. The plot shows the results of a Hartree-Fock calculation (light blue), which largely misses the correlated spin fluctuations of the electrons between atoms, or any long-range antiferromagnetic (AFM) order which is shown to emerge in the exact FCI (red) result. We also show the results of a classically optimized CCSD state (grey), which despite being accurate for on-atom and nearest-neighbor magnetic fluctuations, does not accurately describe the decay of the correlations with distance, exhibiting an erroneously slow decay of this AFM order. After encoding the exact state on a statevector simulator, we can also compare the approaches to sampling this observable as would be done on a quantum device. We can evaluate this spin correlator by sampling the corresponding Pauli operators directly, grouped qubitwise into mutually commuting sets (purple), or by sampling excitation amplitudes from the classical shadow procedure and mapping to the classical CCSD surrogate state as described above (orange). Both used a total of ten sets of 40,960 shots to estimate \toadd{the bias from the CCSD mapping, as well as allowing a faithful comparison of the statistical errors between these two sampling approaches.}

As expected, directly measuring the Pauli strings of the correlation function incurs no systematic error, with the exact results within statistical errors over all length scales. This contrasts with the approach relying on mapping the \toadd{shadow} sampled state to a CCSD model, which exhibits a systematic \toadd{bias} originating from the incompleteness of the ansatz, as well as covariances arising from products of stochastic $t$-amplitudes in the non-linear functional for the RDM of Eq.~\ref{eq:ccrdm}. However, we find that this systematic error is always less than the classically optimized CCSD state, and results in a reduction in the particularly poor description of the long-range correlations by a factor of two, partially correcting the deficiency in the classical CCSD description. Furthermore, we find a large reduction in the statistical errors associated with this estimate compared to the Pauli estimator. 

Whether this approach and the level of systematic error is acceptable compared to the Pauli estimator will be dependent on the system, accuracy required, and the level of stochastic error in the Pauli estimator for a given shot budget. To consider this, and the scaling of the results with system size, in Fig.~\ref{fig:SzSz}(b) we consider the same-atom spin-spin correlations, $\langle \hat{S}_z(H_0)\hat{S}_z(H_0) \rangle$, and decompose the overall error from the CCSD surrogate model \toadd{into its systematic bias and variance, and compare these to the random error of the Pauli estimator for this local expectation value for the same number of shots.} These errors are then compared as the system grows in size as more atoms are included for a fixed interatomic distance. For this correlation function the statistical error in the Pauli estimator grows significantly quicker over the accessible system sizes compared to both the statistical error and systematic error of using the CI amplitudes measured on the quantum device and mapped to the CCSD ansatz. In this case therefore, the systematic error is small enough that it is well inside the larger random errors of the Pauli estimators, clearly demonstrating a benefit in the approach. Furthermore, both the random and systematic errors in the mapped CCSD state appear to grow linearly, compared to the expected higher power of system size for the Pauli estimator.

However, we expect the effectiveness of this approach to be dependent on the correlation function of interest, with Fig.~\ref{fig:SzSz}(a) already demonstrating that the larger distance correlation functions exhibit a larger systematic error. Nevertheless, a systematic improvement in this error can always be found via the inclusion of higher order excitations in the coupled cluster model, which will be considered in the future. \toadd{For further comparison between these approaches, we can combine the bias and variance errors of the estimates into a mean squared error in the spin correlation functions for a given shot budget, which we provide in Appendix~\ref{app:MSE_properties} for further insights into the efficiency of the CCSD mapped shadow approach.}

\section{Application to quantum embedding} \label{sec:embedding}
In this section we consider a further role for the effective combination of quantum excitation amplitude sampling with classically tractable methods; to enable more effective quantum embedding for extending the applicability of quantum computation to larger electronic systems. Quantum embedding methods exploit locality in correlated physics in order to partition a large quantum system into smaller fragments, while admitting a reduced description of the quantum fluctuations between each fragment and its environment as an associated quantum `bath'. The reduced description of the inter-fragment correlations is found via a classical level of theory (generally mean-field) over the entire system. This enables each fragment of the system to be mapped to a auxiliary `cluster' model where the fragment is coupled to its bath space describing these simplified environmental fluctuations. These cluster models can then be solved with a correlated approach, often followed by a self-consistency condition in order to optimize the full system description from the solutions of each fragment. A variety of quantum variables can be used to formulate these theories and describe the electronic structure of each fragment, including their local Green's function (in e.g. dynamical mean-field theory)~\cite{PhysRevX.6.031045,rungger2020dynamicalmeanfieldtheory,selisko2024dynamicalmeanfieldtheory,Keen_2020,PhysRevB.107.165155,Jaderberg_2020,jamet2024andersonimpuritysolverintegrating}, spectral moments~\cite{PhysRevB.103.085131,PhysRevResearch.3.033230,PhysRevB.98.235132}, or density matrices in the case of density matrix embedding theory~\cite{rubin2016hybridclassicalquantumapproachlargescale, Vorwerk2022,DMET_montanaro,D2SC01492K,Cao2023,PhysRevResearch.3.013184,Gujarati2023,PhysRevResearch.3.033230}. These variables are in turn used to define local expectation values, or combined to extract non-local (e.g. energetic) properties of the full system.

These quantum embedding methods are widely expected to be essential for the applicability of quantum computers to larger realistic electronic problems~\cite{DMET_montanaro,rungger2020dynamicalmeanfieldtheory,rubin2016hybridclassicalquantumapproachlargescale,D2SC01492K}. They enable a targeted use of limited quantum information processing resources to correlated local subspaces with relatively weak inter-fragment correlations, and aim to extend beyond the capabilities of classical solvers. Furthermore, the fragmentation of a system into smaller coupled problems enables quantum computers to circumvent the formal `orthogonality catastrophe' arguments which are expected to limit the applicability of quantum computers for eigenstate preparation of larger systems~\cite{simon2025dividingconqueringvanvleck,PRXLife.3.013003}.
The established quantum embedding techniques, relying on low-body Green's functions or reduced density matrices as descriptions of the correlated fragments, have a substantial drawback when combined with quantum solvers; a simple reconstruction a full system description from these local variables results in a loss of $N$-representability, most clearly manifesting in (among other things) an incorrect total electron number for the system~\cite{Nusspickel2023_reconstruction}. While for lattice models electron number can be imposed exactly as a local symmetry constraint in the solver, this is not true for {\em ab initio} systems where the correlations can rearrange the electron density between fragments and no conserved local electron number can be defined. Therefore, a global self-consistency in order to optimize a chemical potential and ensure correct total electron numbers amongst the combined fragmented models is essential for the method. However, both this electron number and other self-consistencies are numerically sensitive to the probablistic or other (e.g. trotterization) errors in quantum algorithms, and found to limit the efficacy of these quantum embedding techniques in conjunction with quantum solvers~\cite{Keen_2020,iijima2023accuratequantumchemicalcalculations}.

Recently, it was proposed to instead use the configurational amplitudes themselves as the quantum descriptors of the correlated physics in each cluster model~\cite{Nusspickel_systematic}. This enables the local correlated quantum fluctuations of each cluster model to be recombined into a description of the excitation amplitudes of the full system Fermi vacuum. This has the advantage of ensuring an implicit wave function description over the full system at all stages, rigorously enforcing representability of the description, while symmetries such as global electron numbers are conserved by construction. Combined with a density matrix embedding theory approach to constructing compact bath orbital expansions~\cite{Nusspickel_systematic,doi:https://doi.org/10.1002/9781119129271.ch8,doi:10.1021/acs.jctc.6b00316,PhysRevB.101.045126,PhysRevB.102.165107}, this substantially simplifies the quantum embedding and ensures robustness to noise and probablistic solvers in the extraction of non-local and energetic quantities over the system~\cite{Nusspickel2023_reconstruction}. 

Specifically, we can then recombine the single and double excitation amplitudes within each cluster model, $c_{ij}^{ab,X}$, where $X$ labels the individual cluster models, into approximate excitation amplitudes over the whole system. This can be done, while rigorously removing the double-counting arising due to the bath spaces of one cluster overlapping with a different fragment space, with more details on this construction found in Refs.~\cite{Nusspickel_systematic,Nusspickel2023_reconstruction}. These excitation amplitudes of the complete system, approximated by the locality assumptions of the embedding (which can be systematically relaxed with larger bath spaces~\cite{Nusspickel_systematic}), can then be used to extract the total energy of the system via the mixed estimator of Eq.~\ref{eq:pert_energy}, or properties via Sec.~\ref{sec:properties}. This removes the necessity for multiple calculations per fragment required for self-consistency over electron numbers or other matching conditions. Furthermore, it has been shown to dramatically improve energies and non-local expectation values compared to recombining cluster density matrices~\cite{Nusspickel2023_reconstruction}. These properties are particularly appealing in the use of quantum algorithms as cluster solvers.

\begin{figure}[t]
    \centering
    \includegraphics[width=.5\textwidth]{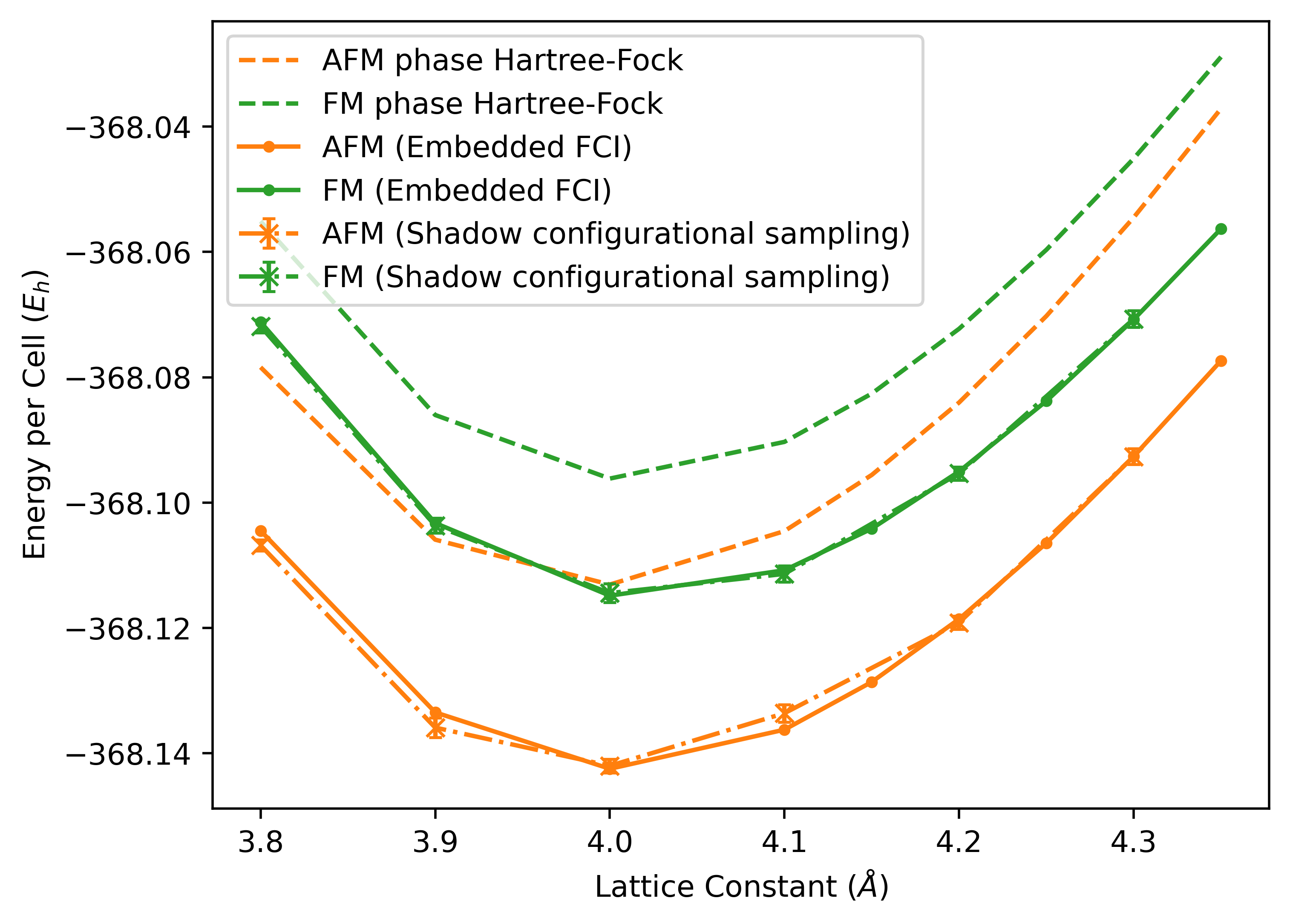}
    \caption{{\bf Smooth and statistically resolved equations of state for Nickel Oxide phases, with shadow sampling of quantum solvers for each fragment demonstrating robustness of quantum embedding approach.} Equation of state for the ferromagnetic and antiferromagnetic phases of NiO, employing an excitation amplitude based quantum embedding to fragment the four-atom unit cell into ten fragments. Compared are exact solutions for each cluster (embedded FCI) and a classical shadow sampling procedure for the excitation amplitudes (shadow configurational sampling) from the exact statevector. Each cluster solution was sampled with 10,000 shots, repeated 12 times to estimate the mean and error of the total energy. Also included is the Hartree--Fock energy for comparison. A $2\times2\times2$ $k$-point grid was used, with the gth-szv-molopt-sr basis and associated gth-pade pseudopotential.}
    \label{fig:NiO_fci}
\end{figure}

Quantum configurational sampling via classical shadows and the mixed energy estimator therefore naturally and seamlessly combine with quantum embedding, partitioning a large quantum system into more tractable fragments in a robust fashion. We consider the application to a 32-atom rocksalt supercell of the archetypal binary transitional metal oxide, Nickel Oxide, a paradigmatic strongly correlated material. By using a statevector simulation for the solution of the individual fragmented cluster models, we demonstrate the applicability of this quantum embedding and classical shadow framework. We consider both an antiferromagnetic (AFM) and ferromagnetic (FM) phase, by converging the low-spin and high-spin states respectively with unrestricted Hartree--Fock starting from a spin symmetry broken solution and converging to static moments on the Nickel atoms as desired using the {\tt pyscf} code~\cite{PySCF2017,PySCF2020}. The particular {\mbox Type-II} antiferromagnetic order consists of planes of spin-up and spin-down polarized Ni$^{2+}$ ions alternating along the $\langle 111 \rangle$ direction of the conventional cell. Neutron diffraction experiments have shown this to be the ground state order~\cite{PhysRev.110.1333}, with a second-order transition to paramagnetic order at a N\'eel temperature of 524~K~\cite{PhysRevB.28.6542,PhysRevB.79.172403}. Accurate computation of the spin gap between these phases and its variation with cell volume is crucial to understand the nature of these phase transitions, its coupling to structural distortions, and the effective Heisenberg exchange interaction of these materials~\cite{PhysRevB.93.195137,10.1063/5.0114080}.

In Fig.~\ref{fig:NiO_fci}, we consider the energy of these NiO phases across a changing lattice constant in the rocksalt cells. We fragment the four-atom primitive cell into ten fragments of intrinsic atomic orbitals~\cite{Knizia2013}. These are grouped into the valence orbitals on each oxygen, and the Nickel atom orbitals further partitioned into disjoint sets as $(3s, 3p)$, $(4s, 4p)$, $(3d_{xz}, 3d_{xy}, 3d_{yz})$, which mix to give the $t_{2g}$ crystal field orbitals, and $(3d_{x^2-y^2}, 3d_{z^2})$ which mix for the $e_g$ set. Combined with the projected (symmetry-broken) bath states from DMET to describe the fluctuations from these fragments to the wider system~\cite{Cui2020_ab_initio_DMET, Nusspickel_systematic}, the clusters have a maximum of 16 spin-orbitals in total, constructed using the {\tt vayesta} code. These are exactly solved for their ground state in a statevector simulation, and random Clifford operations applied in order to sample 10,000 shots to build the shadow state for each cluster. Each cluster estimate of the excitation amplitudes are then combined, assuming translational symmetry of the primitive cell, in order to compute the mixed energy estimate. This is repeated twelve times in order to achieve an error bar on the energy estimate for each lattice constant. 

It can be seen in Fig.~\ref{fig:NiO_fci} that these estimates agree well within the error bars of the exact cluster amplitudes obtained from FCI. Furthermore, the agreement demonstrates the robustness of the estimate to shot noise and advantages derived from only needing to perform a single optimization and stochastic sampling of each cluster, avoiding chemical potential and other self-consistent optimization loops. We find that the inclusion of the local correlation from the embedding procedure significantly increases the spin gap of the NiO as expected, and increases the equilibrium cell volume towards the experimental results. More care however will be required to ensure convergence with respect to $k$-points, basis set and bath sizes before a rigorous comparison with experimental results or other theoretical studies is possible~\cite{Cao2023,Cui2020_ab_initio_DMET}. However, this at least demonstrates that this novel shadow sampling combined with quantum embedding protocol is able to resolve correlation driven structural and spin-gap changes for realistic correlated transition metal materials in a robust fashion.

\begin{figure}[t]
    \centering
    \includegraphics[width=.5\textwidth]{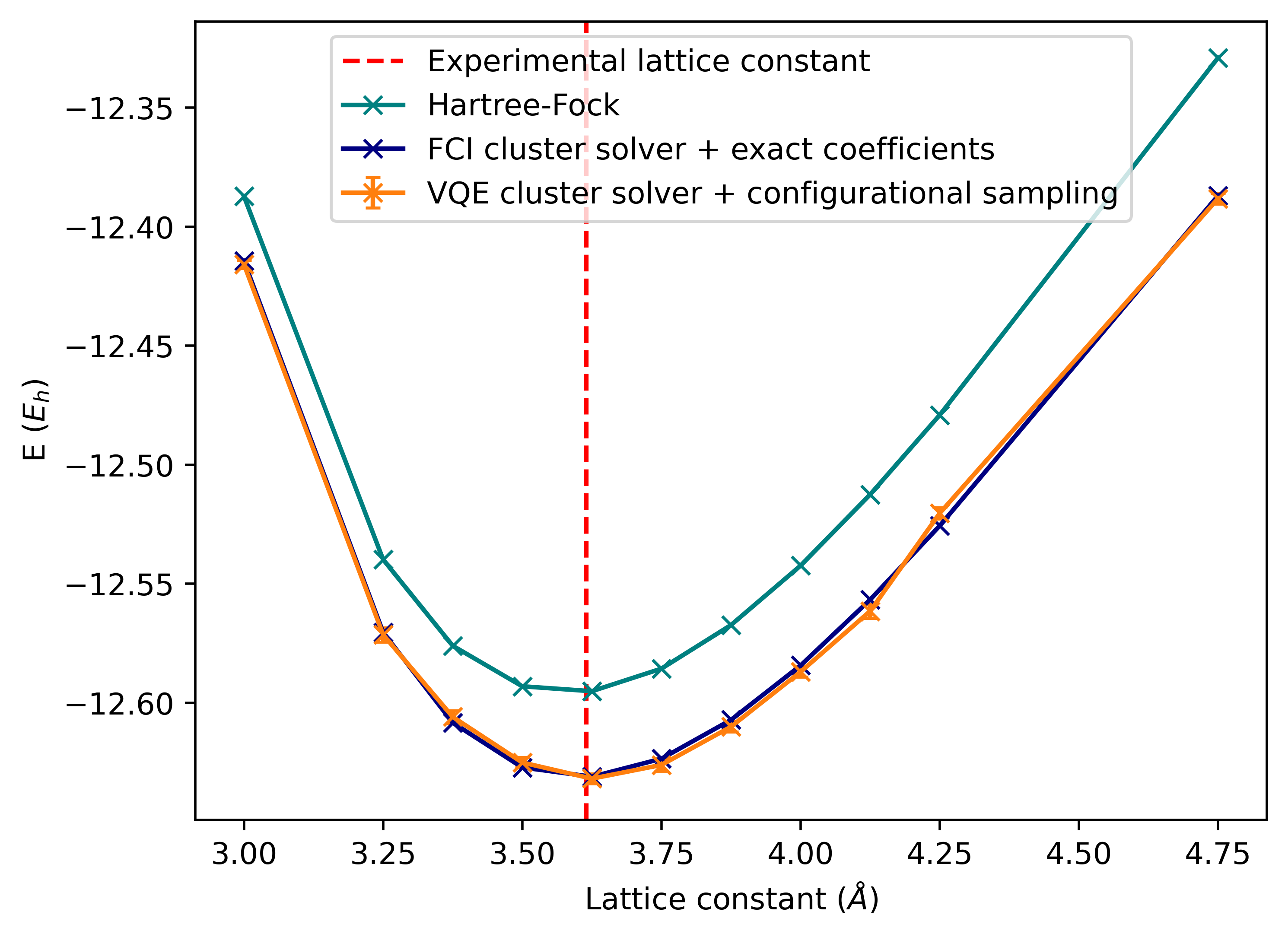}
    \caption{{\bf Equation of state of $3\times3\times3$ cubic-BN, fragmenting the system into atomic valence orbital sets with VQE solvers.} Each resulting cluster (with a maximum of 12 qubits) was solved with a UCCSD ansatz, before the excitation amplitudes sampled with 4,000 shots to obtain a total energy at each cell size. This was repeated 12 times to estimate the mean and error of the total energy. Also compared is a FCI solution to the same quantum embedding, giving excellent agreement, and the Hartree--Fock result for the system. A gth-dzvp basis was used with gth-pade pseudopotential.}
    \label{fig:BN_vqe}
\end{figure}

We can also go beyond exact statevector cluster solutions, and consider a full NISQ workflow using VQE as the cluster solver, followed by the shadow tomography of the excitation amplitudes of each cluster and recombining for the full system energy. In Fig.~\ref{fig:BN_vqe} we apply this to cubic Boron Nitride in an FCC lattice (space group $F{\overline 4}3m$), considering atomic fragments consisting of the $2p$ orbitals on each atom, giving a cluster size of 12 spin-orbitals, and the core $(1s, 2s)$ orbitals of each atom in the primitive cell treated separately. Within a $3\times3\times3$ $k$-point sampled cell, each cluster was solved with VQE with the unitary coupled-cluster (UCCSD) ansatz. Only one optimization per cluster was required, with 4,000 shots used to build a classical shadow and extract out the full system mixed energy estimator. Twelve independent samples of the shadow state were taken to obtain a mean and variance on the energy estimate, with error bars barely visible on the plot. The equivalent FCI cluster results are always within two standard deviation and the error can be systematically reduced with more samples and improved ground state preparation algorithms. We find that the effect of the correlated treatment is to enlarge the lattice parameter to coincide with the experimental result of $c=3.615$\AA~\cite{D4TC00039K}. The correlations in this system are relatively modest, yet the reliability and robustness of the overall protocol remains, allowing smooth equations of state even when combined with approximate state preparation quantum algorithms. This allows us to go beyond the largely single-geometry energy estimates of previous approaches in combining embedding approaches with quantum solvers.


\section{conclusion}
This work demonstrates a broad scope of advances in simulating {\em ab initio} electronic structure on quantum devices which can be enabled via the efficient sampling of a limited set of configurational amplitudes via shadow tomographic techniques. By relying on a non-vanishing projection onto a classically preparable state, we show that a non-variational mixed energy estimator that is linear in the statevector can directly target the correlation energy. Away from highly correlated limits, this demonstrates a substantially reduced variance in the energy expectation value for realistic systems, as well as its scaling with system size, dramatically reducing shot numbers required for energy estimation. The reliance on a classical state with appreciable overlap and that can be prepared efficiently on a quantum device means that asymptotic scalings must be exponential (akin to state preparation protocols more generally). However, we find numerically that this limitation is far enough away as to expect a substantial window of algorithmic advantage in the approach for reducing sampling overheads of prepared quantum states for near-term devices. Furthermore, there is scope for improved classical states to be used for larger or more correlated systems, including tensor networks, or self-consistent refinement of a single determinant state~\cite{SCHAFER19711}, as well as optimization of the shadow sampling protocols~\cite{Hadfield2022,PRXQuantum.4.040328}.

We find the performance of this energy estimate to be particularly resilient to noise, constructing a noise model interpolated and validated from IBMQ simulations. The non-variationality of the method allows for correlation energies to be found from a VQE-optimized state beyond noise levels where direct sampling of Pauli strings qualitatively fails. The same sampled configurational amplitudes can also be used to directly map to a surrogate coupled cluster model, from which other non-energetic properties can be efficiently extracted on classical computing resources. This is found to exhibit a clear quantum enhancement of correlation functions of correlated systems over the direct optimization of the classical surrogate for the same system, and in many cases the scaling with system size and systematic errors are preferable over direct sampling methods of these observables.

Finally, we consider how tomography over these configurational amplitudes enables application of an efficient quantum embedding framework for application to larger systems. Working with configurational amplitudes obviates the requirement of self-consistent loops, which have previously been found to be particularly sensitive to numerical noise in quantum solvers in alternate quantum embedding frameworks. This enhanced robustness enables smooth equations of state to be found in correlated materials, as applied to the geometry-dependence of the spin gap of Nickel Oxide and Boron Nitride. While convergence with respect to all technical parameters required for predictive accuracy compared to experimental results is still out of reach for current quantum solvers, the combination of shadow tomography and quantum embedding outlines an effective new route to accelerate the use of quantum solvers in large scale {\em ab initio} materials science workflows.


\section*{Acknowledgments}

The authors wish to thank Mohammadreza Mousavi for helpful comments over the course of this work.
This research has been partially supported by the EPSRC project on Verified Simulation for Large Quantum Systems (VSL-Q), grant reference EP/Y005244/1 and the EPSRC project on Robust and Reliable Quantum Computing (RoaRQ), Investigation 009 Model-based monitoring and calibration of quantum computations (ModeMCQ), grant reference EP/W032635/1. For the purpose of open access, the author has applied a Creative Commons Attribution (CC BY) license to any Author Accepted Manuscript version arising. The data that support the findings of this article are openly available~\cite{datarepo}. We acknowledge the use of IBM Quantum services for this work. The views expressed are those of the authors, and do not reflect the official policy or position of IBM or the IBM Quantum team.


\appendix

\section{Mixed energy error analysis}\label{app:full_error}
In the main text we presented an error scaling for the mixed estimator ignoring the effect of the overlap with the reference state ($c_0$). We can incorporate this effect using the typical formulae for the variance of the distribution for a function composed of the sums and ratios of random variables. 
\begin{align}
    f = aA + bB \implies \sigma_f^2 = a^2 \sigma_A^2 + b^2 \sigma_B^2 \\
    g = \frac{A}{B} \implies \frac{\sigma_g^2}{\mathbb{E}[g]^2} = \frac{\sigma_A^2}{{\mathbb{E}[A]^2}} + \frac{\sigma_B^2}{{\mathbb{E}[B]^2}}
\end{align}
Indeed, again ignoring covariances between the distributions of the $c_{ij}^{ab}$ and $c_0$ amplitudes, the total variance in the correlation energy will be given by
\begin{align}
    \epsilon^2 = \frac{1}{M}&\frac{\epsilon_{\text{amp}}^2}{|c_0|^2} \left(\sum_{ijab}|2v_{ijab}-v_{ijba}|^2 + \right. \nonumber\\
    &\left. \frac{1}{|c_0|^2} \left(\sum_{ijab}(2v_{ijab}-v_{ijba})c_{ij}^{ab}\right)^2 \right) .
\end{align}
This highlights the formal dependence of the random error on $|\langle \Phi|\Psi\rangle|^{-2} \sim 1/c_0^2$, which returns to the previous shot scaling estimate of Eq.~\ref{eq:shotsmixed} in the case that $c_0=1$ with no uncertainty in this estimate. This analysis also assumes that all excitation amplitudes have the same variance of $\epsilon_{\text{amp}}$ -- a reasonable assumption given the unbiased selection of Clifford circuits in the shadow state construction. This variance will scale as $\log(N)$, as shown in Eq.~\ref{eq:errorperamp}.

Since the $c_{ij}^{ab}$ and $c_0$ amplitudes are all calculated from the same set of measurements used to form the classical shadow there will also be covariances between these estimates. These covariances will decrease as $1/M$ and so will not changes the overall dependence of the variance of the distribution on the shot count. These covariances between $c_0$ in the denominator and the $c_{ij}^{ab}$ in the numerator can however lead to the distribution of the estimate of the correlation energy at a fixed shot count to have a systematic error. This error can be removed by using a separate set of shots to estimate uncorrelated shadow states for $c_0$ and the $c_{ij}^{ab}$ amplitudes, however, in practice this error seems to be much smaller than the standard deviation. This is also validated by previous literature highlighting the small covariances in the distribution for configurational amplitudes~\cite{Kiser_2024}.

\section{Hamiltonian dependent variance factor} \label{app:ham_coeffs}
\begin{figure}[H]
    \centering
    \includegraphics[width=0.5\textwidth]{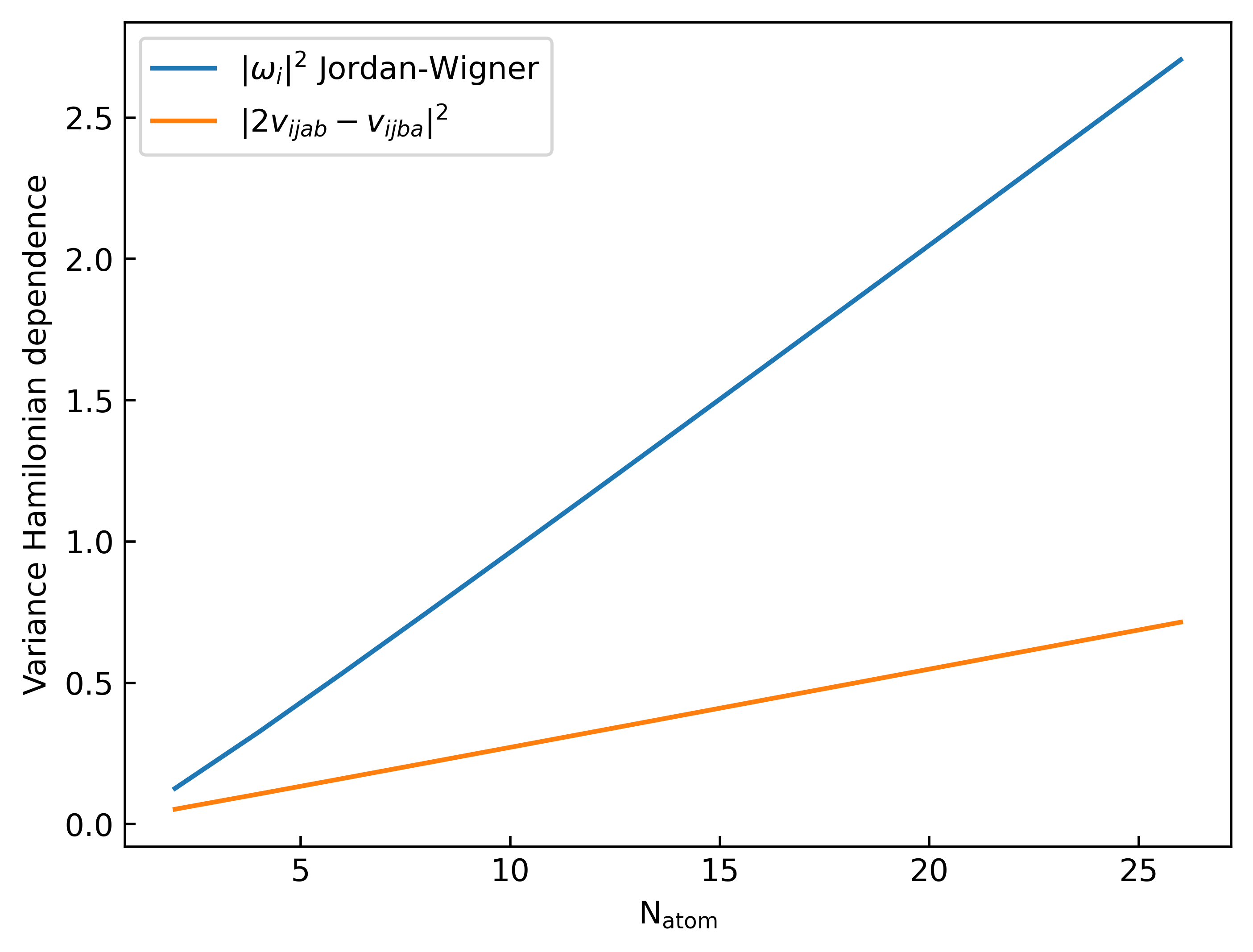}
    \caption{\textbf{Hamiltonian dependent scaling of the variance prefactors.} The Hamiltonian dependent factor in the variance expression for the Pauli (blue) and mixed (orange) estimators for a symmetrically stretched STO-3G basis Hydrogen chain with 1.5$\AA$ interatomic distance.}
    \label{fig:hamiltonian_term_scaling}
\end{figure}

The scaling of the shot number to achieve a desired precision for both the Pauli and mixed estimators depend on different prefactors which are functions of the Hamiltonian of interest (see Eqs.~\ref{eq:shotspure} and \ref{eq:shotsmixed} respectively).  For the direct Pauli sampling, the variance of the energy estimate depends on the factor
\begin{equation}\label{eq:pauli_hamiltonian_term}
    \sum_{a}^{\mathcal{P}}w_a^2 \mathrm{Var}[\hat{P}_a],
\end{equation}
where the variance of the Pauli string, $\hat{P}_a$, is bounded $0<\mathrm{Var}[\hat{P}_a]<1$. 
For the mixed correlation energy estimator, the scaling in shot number instead depends on the difference of electron repulsion integrals, $v_{ijab}$, as
\begin{equation}\label{eq:mixed_hamiltonian_term}
    \sum_{ijab}|2v_{ijab}-v_{ijba}|^2.
\end{equation}

To numerically compare these quantities and investigate how they depend on system size, in Fig.~\ref{fig:hamiltonian_term_scaling} we plot a comparison of the terms in Eqs.\ref{eq:pauli_hamiltonian_term} and \ref{eq:mixed_hamiltonian_term} for increasing length of hydrogen chains in a minimal basis with an atomic spacing of $1.5$\AA~in a canonical basis representation. The variance of each Pauli string in Eq.~\ref{eq:pauli_hamiltonian_term} is unknown and dependent on the state of interest. We therefore assume that the $\mathrm{Var}[\hat{P}_a]=1$ for all Pauli strings, and take the weights of the Hamiltonian, $w_a$, to arise from the Jordan-Wigner fermion-to-qubit mapping. With this simplifying assumption, the prefactors are both found to increase linearly with system size, albeit with a substantially smaller gradient for the mixed estimator prefactor. This validates the arguments of the main text that the correlation energy estimator has a smaller prefactor, encoding just the electron repulsion terms rather than all the physics of the Hamiltonian.
We also note this is just the prefactor scaling, with the full scaling in shot number for the Pauli estimator then multiplied by a polynomial factor, while for the mixed estimator the explicit system size dependent factor is logarithmic, combining to result in the favorable scaling observed in 
Fig.~\ref{fig:hstretch_errors}(b).

\section{Mean squared errors of magnetic estimates of Hydrogen chains}\label{app:MSE_properties}

\begin{figure*}[ht!]
    \centering
    \begin{subfigure}[t]{0.5\textwidth}
        \centering
        \includegraphics[height=6.5cm]{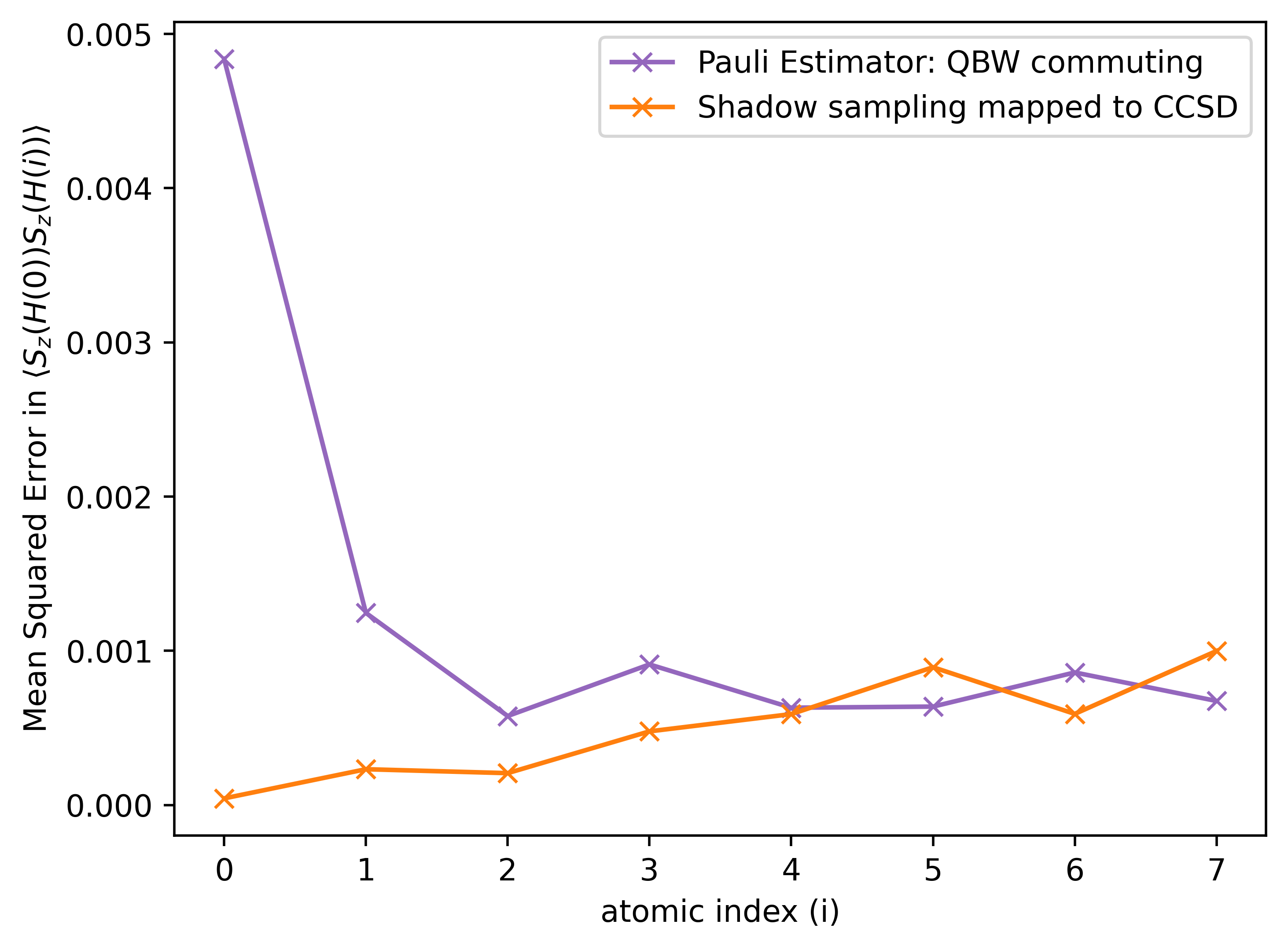}
    \end{subfigure}%
    ~
    \begin{subfigure}[t]{0.5\textwidth}
        \centering
        \includegraphics[height=6.5cm]{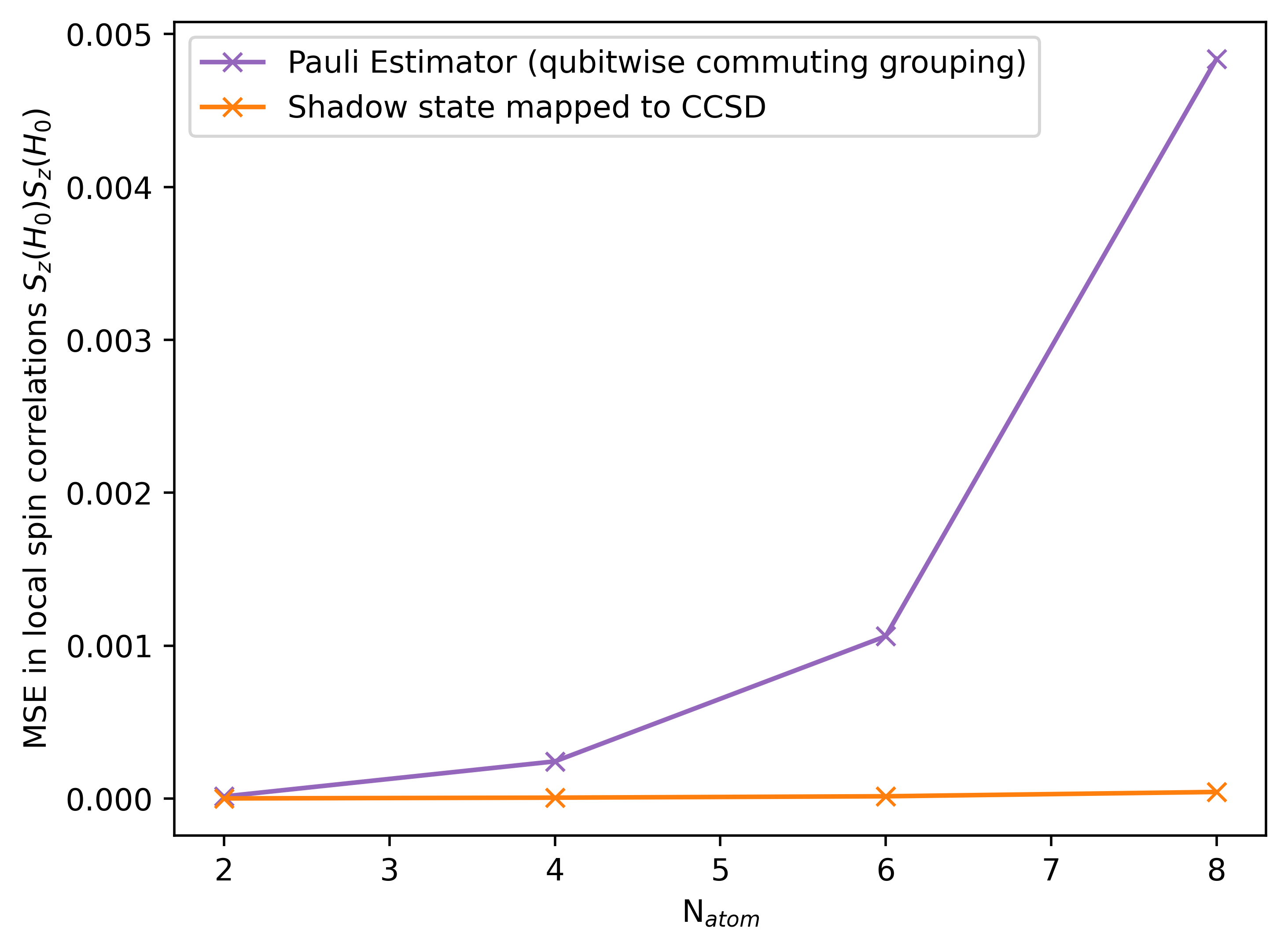}
    \end{subfigure}%
    \caption{\toadd{{\bf Mean squared errors (MSE) in the magnetic correlations of Hydrogen chains} Left: The MSE spin-spin correlation errors for an eight atom hydrogen chain ($1.5$\AA~ symmetric stretching) between the first atom and its $i^\text{th}$ nearest neighbor in a STO-3G basis. Ten sets of 40,960 shots are used to calculate the MSE from the exact statevector, comparing a direct sampling of qubitwise grouped Pauli strings (purple) and a mapping of the shadow state to a CCSD ansatz (orange). Right: The RMS error in the local (same-atom) spin-spin correlation function sampled over 40,960 shots for an N$_{\mathrm{atom}}$ hydrogen chain of increasing length.}}
    \label{fig:SzSz_app}
\end{figure*}

\toadd{To further characterize the errors from the property estimates of Sec.~\ref{sec:properties}, we can consider the mean squared errors (MSE) in the magnetic correlation functions of Fig.~\ref{fig:SzSz} in the main text. This allows us to combine the bias and variance contribution to the errors of the shadow state mapped to the coupled cluster form, and compare directly to the corresponding MSE of the Pauli estimator of these properties for a given shot budget.
In Fig.~\ref{fig:SzSz_app}, we show these MSE errors in the two-point magnetic correlations, both between increasingly distant atom pairs for a fixed chain length (left plot), and for the same-atom magnetic correlations but with an increasing length of the overall chain. Both estimators sampled the same number of shots (40,960) from the exact ground statevector.}

\toadd{We find that the MSE errors in these magnetic correlations increase approximately linearly with the distance between the atomic sites considered, with RMS errors equivalent between the two approaches from a distance of approximately four sites. This is largely driven from the increasing bias in the CCSD ansatz, also observed in Fig.~\ref{fig:SzSz}. However, this bias is minimal for same-spin or close neighbour correlation functions, where significant advantages over the Pauli sampling are found. This advantage is only compounded for increasing system sizes (right plot), where we find an RMS error which is largely independent of system size, in contrast to the rapidly increasing MSE error of the Pauli expectation (driven by the increasing variance of this estimator).}

\section{Calculation of stabilizer overlaps}\label{app:stabilizer_calc}
An important part of the evaluation of configurational amplitudes from the classical shadow is the calculation of the overlap between two stabilizer states~\cite{Huggins_unbiasingAFQMC}, which arises when evaluating the expectation value of the operators with the classical shadow density matrix as found in Eq.~\ref{eq:ampexpectation}, as
\begin{equation}
    \braket{\phi_i|\psi} = 2(2^N+1)\frac{1}{N_S}\sum_{j=1}^{N_S}\braket{\phi_i|U_j^{\dagger}|b_j}\braket{b_j|U_j|0}.
\end{equation}
In this work, we are restricted to considering $\ket{\phi_i}$ (as well as $\ket{b_j}$) to be computational basis states, with $U_j$ an $N$-qubit Clifford unitary, amounting to the calculation of the overlap (including its phase) between a stabilizer state and a computational basis state.
An approach for this calculation is given in Ref.~\cite{geometry_of_stabilizers}, however, given the simplification of only requiring overlap with a computational basis state we can simplify the algorithm to avoid tracking of the global phase of the stabilizer state. 

To perform this calculation we first reduce the stabilizer matrix to a canonical form~\cite{geometry_of_stabilizers}. The absolute value of the inner product of the stabilizer state with each computational basis state is given by $2^{-s/2}$ where $s$ is the $X$-rank of the reduced stabilizer matrix (the number of generators containing an X or Y Pauli)~\cite{geometry_of_stabilizers}. To calculate the phase we then perform a measurement of the circuit which creates the stabilizer state $U_j^{\dagger}\ket{b_j}$ using the algorithm shown in ref.~\cite{PhysRevA.70.052328}. This results in one of the computational basis states which overlap with the stabilizer state $U_j^{\dagger}\ket{b_j}$. From this measured state we can use the generators in the canonical stabilizer matrix to transform it into each of the basis states of interest, i.e. $\ket{0}$ and $\ket{\phi_i}$, while tracking the phase generated by the action of each generator on the state. The Gaussian elimination step scales as $\mathcal{O}(N^3)$, which is the most costly step of the algorithm and it must be repeated for each measurement in the shadow $U_i^{\dagger}\ket{b_i}\bra{b_i}U_i$. The overall cost of the post processing will therefore be determined by the number of basis states that the phase needs to be calculated for, resulting in an overall scaling in order to compute the singles and doubles excitation amplitudes of $\mathcal{O}(M(N^4 + N^3))$.

We can give an explicit example of calculating 
$\braket{\phi_i|U_j^{\dagger}|b_j}\braket{b_j|U_j|0}$ for the $j$th element of the classical shadow. Assuming a measured state $\ket{b_j} = \ket{1011}$ and Clifford unitary with stabilizers $[-ZYYX, +ZXXI, -ZIXZ, -YYXX]$ the stabilizer state $U_j\ket{b_j}$ has a stabilizer matrix given by:
\begin{align}
&[X, Z, I, Y, -1]\nonumber\\
&[Y, X, Y, Y, 1]\nonumber\\
&[X, I, X, Y, 1]\nonumber\\
&[Y, Z, I, Z, -1]\nonumber
\end{align}
which after the Gaussian elimination procedure becomes the reduced stabilizer matrix:
\begin{align}
&[Y, Z, I, Z, -1]\nonumber\\
&[Z, X, Z, I, -1]\nonumber\\
&[I, Z, X, I, -1]\nonumber\\
&[Z, I, I, X, -1].\nonumber
\end{align}
We then perform a measurement and obtain the computational basis state $\ket{0000}$, from which we aim to compute the overlaps with basis states $\ket{0000}$ and $\ket{0101}$. Since we measured the basis state $\ket{0000}$ we only need to work out the phase of the overlap with $\ket{0101}$. Starting from the measured state $\ket{0000}$ we can apply the stabilizer generator
$[Z, X, Z, I, -1]$, resulting in the state $\ket{0100}$ with a phase -1, and then apply 
$[Z, I, I, X, -1]$ which results in the state $\ket{0101}$ with a phase +1. Since the $X$-rank of the reduced matrix is four, the overall value of $\braket{\phi_i|U_j^{\dagger}|b_j}\braket{b_j|U_j|0}$ is $2^{-4}$.


\section{Direct configurational sampling} \label{app:directconfigsampling}

\begin{figure*}
    \centering
    \begin{subfigure}[H]{0.5\textwidth}
        \centering
        \includegraphics[height=6.5cm]{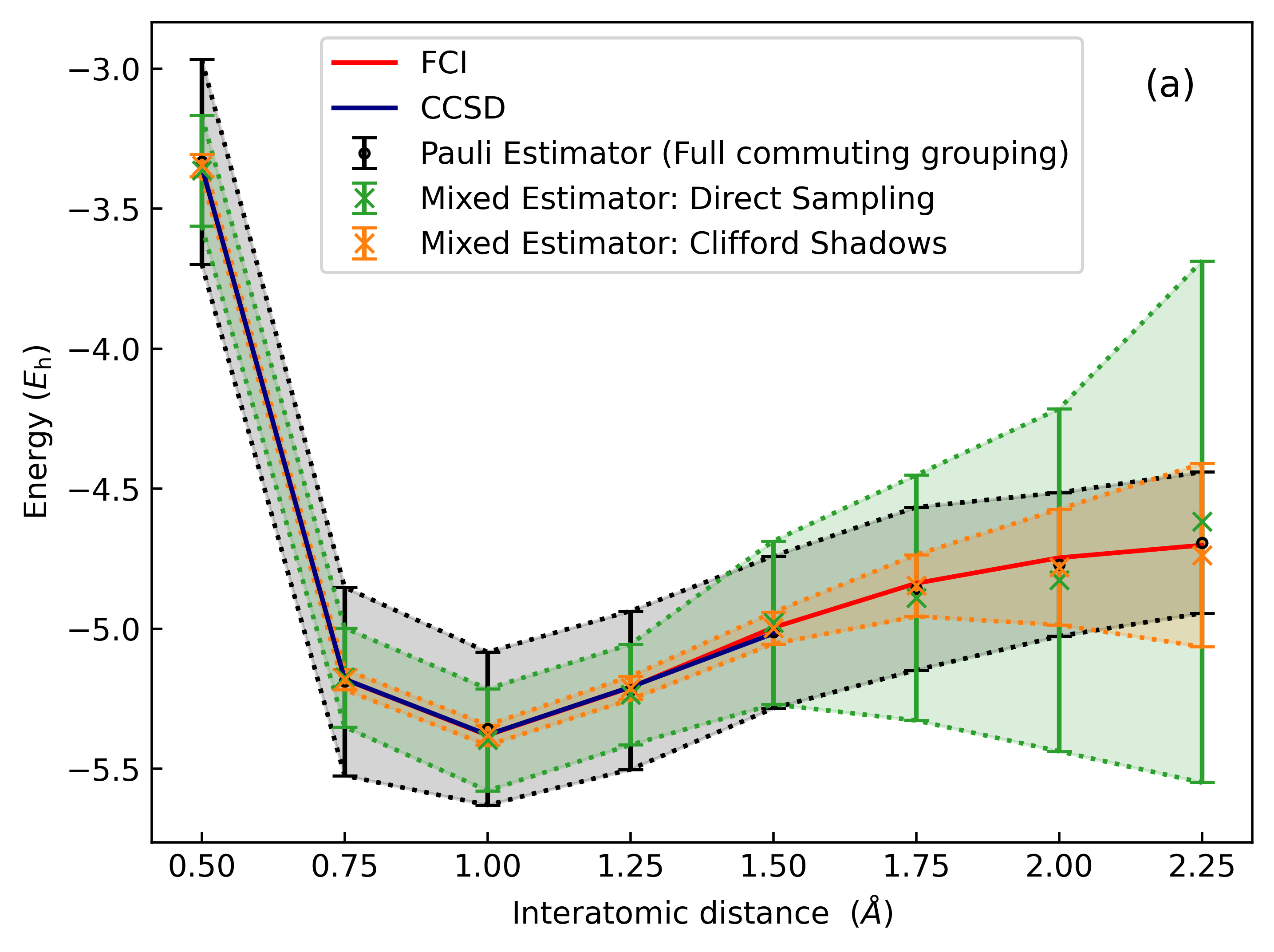}
       
    \end{subfigure}%
    ~
    \begin{subfigure}[H]{0.5\textwidth}
        \centering
        \includegraphics[height=6.5cm]{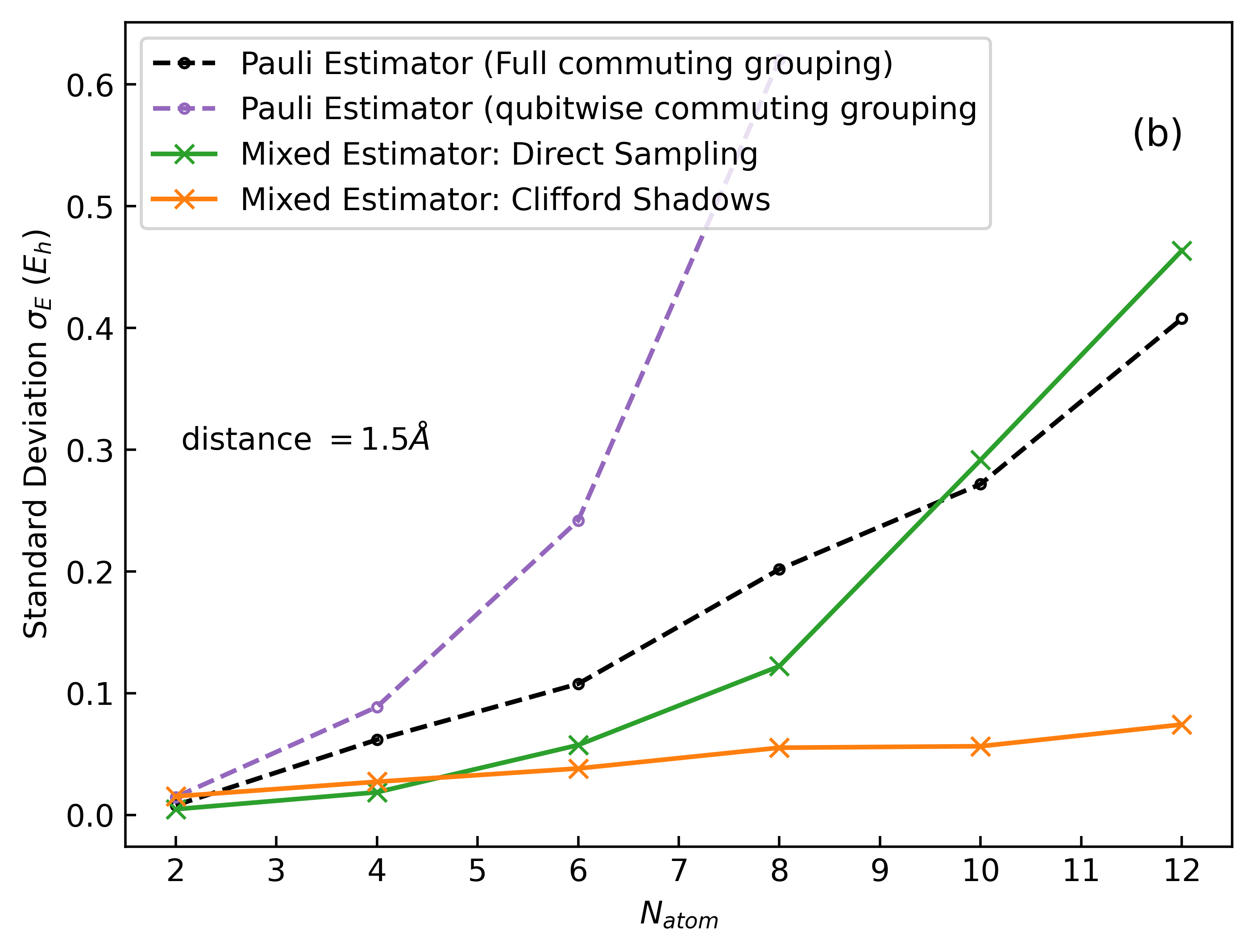}
        
        \label{fig:N_atom_vs_error}
    \end{subfigure}%
    \caption{\textbf{Variance scaling for mixed and Pauli estimators of Hydrogen chains over changing correlation strengths and chain length, analogous to Fig.~\ref{fig:hstretch_errors}, but including a direct sampling of excitation amplitudes in the mixed estimator.} (a) Comparison of the distributions of the Pauli and mixed energy estimators, comparing classical shadows to the direct evaluation of the excitation amplitudes via diagonalization and measurement of the operators in Eq.~\ref{eq:direct_overlap_op} (green), for the ten atom STO-3G basis Hydrogen chain across a range of interatomic distances. The width of the error bar is the standard deviation of the distribution of energies from $1000$ measurements of the exact statevector. (b) The dependence of the standard deviation of each energy estimator with system size, for a fixed atomic distance of $1.5$\AA and $1000$ shots. The black line again shows the Pauli operators grouped into as few commuting sets as possible, while purple is the same for grouping Pauli operators based on qubit wise commutivity. The orange and green lines represent the standards deviations of the mixed estimator using the direct and Clifford shadows methods of calculating the $c_{ij}^{ab}$ amplitudes.\label{app:direct_sampling}}
\end{figure*}

The configurational amplitudes required to evaluate the mixed energy estimator are found by evaluating operators of the type $\ket{0}\bra{\phi_{ij}^{ab}} + \ket{\phi_{ij}^{ab}}\bra{0}$. While this was done within a classical shadow protocol to efficiently evaluate the polynomially increasing number of expectation values with a logarithmically increasing number of shots, these amplitudes can also be measured directly by evaluating the operators 
\begin{equation}\label{eq:direct_overlap_op}
    \ket{\phi_0}\bra{\phi_{ij}^{ab}} + \ket{\phi_{ij}^{ab}}\bra{\phi_0},
\end{equation}
where $\ket{\phi_0}$ is the Hartree-Fock state and $\ket{\phi_{ij}^{ab}}$ is a doubly excited state. We use this alternative to $\ket{0}\bra{\phi_{ij}^{ab}} + \ket{\phi_{ij}^{ab}}\bra{0}$ to keep the circuit which diagonalizes the operators constant depth with respect to the system size, as $\ket{\phi_0}$ and $\ket{\phi_{ij}^{ab}}$ differ by at most only four qubits (given by the indices $i, j, a$ and $b$) independent of system size. We can then diagonalize and measure each of these operators individually to obtain the expected value of $c_0c_{ij}^{ab}$ and measure $c_0$ separately. We can compare this direct sampling approach to the shadow state tomography to obtain the excitation amplitudes required for the mixed estimator, to assess the efficiency benefit from the shadow state protocol.

In Fig.~\ref{app:direct_sampling}(a) we plot the mean and standard deviation of energy estimates for a Hydrogen chain with 10 atoms at varying interatomic distance with 1000 shots, analogous to Fig.~\ref{fig:hstretch_errors}(a) of the main text, but now including a direct sampling comparison for the excitation amplitudes used in the mixed energy estimator.
The red and navy lines show the result of FCI and CCSD calculations, while the orange and black lines represent the mixed estimator with classical shadow calculation of the overlap and the Pauli estimator of the expectation value with full commuting groups, as previously provided in the main text. 
In this plot however, we also provide the mixed energy estimate where the required amplitudes are directly sampled from the operators in Eq.~\ref{eq:direct_overlap_op} (green line). Furthermore, a classical CCSD calculation is used to optimally distribute the shots over the operators, quasi-optimally reducing the variance in this approach. We can see that this direct sampling approach to the mixed estimator substantially increases the variance compared to shadow tomography for the required amplitudes over all correlation strengths, even if for some geometries the variance is still below that of the Pauli sampled estimator.

In Fig.~\ref{app:direct_sampling}(b) we also show the standard deviation for these different energy estimates using 1000 shots at different system sizes with a fixed interatomic distance of $1.5$\AA. The combination of an additional polynomial factor that arises from needing to split the shot budget between an increasing number of observables and the increased dependence on $c_0$ arising from calculating $c_0c_{ij}^{ab}$ factors leads to a substantially worse scaling with system size compared to the classical shadow approach in almost all scenarios. This validates the benefits found from the combination of shadow sampling and the mixed estimator presented in the main text.

\end{document}